%% file: psaem.tex
\documentclass[journal,10pt]{IEEEtran}
\ifCLASSINFOpdf
\else
\fi
\usepackage{amsmath,mathtools,amsthm,bbm}
\usepackage{algorithm}
\usepackage{algpseudocode}
\usepackage{lindsten,abbreviations}
\usepackage{color}
\usepackage[capitalize,noabbrev]{cleveref}
\usepackage[font=footnotesize]{caption}
\usepackage[font=footnotesize]{subcaption}
\usepackage{enumerate}
\usepackage{booktabs}
\usepackage{url}
\usepackage{tikz}
\usepackage{pgfplots}
\usepackage{grffile}
\usetikzlibrary{calc,positioning,arrows,shadows,arrows.meta}
\pgfplotsset{compat=newest}
\pgfplotsset{every tick label/.append style={font=\tiny}, y tick label style={rotate=90}}
\pgfplotsset{every axis label/.append style={font=\footnotesize}}
\tikzset{block/.style={
		fill=green!40!black!20!white,
		rectangle,
		align=left,
		text width=5.8cm,rounded corners,drop shadow=black!50!white,
		font=\footnotesize}}

\usepackage{pifont}
\newcommand{\cmark}{\ding{51}}%
\newcommand{\xmark}{\ding{55}}%

\newcommand{\myd}{{\rm d}}
\renewcommand{\mid}{\,\vert\,}

\newcommand{\algfontsize}{\small}

\newcommand{\T}{T}
\newcommand{\thml}{{\widehat\theta}}  
\newcommand{\etaml}{{\widehat\eta}}  
\newcommand{\X}{x_{0:\T}}
\newcommand{\Y}{y_{1:\T}}
\newcommand{\Qf}{Q^{\text{Fish}}}
\newcommand{\Qb}{Q^{\text{Bay}}}
\newcommand{\Qaf}{\widehat{Q}^{\text{Fish}}}

\newcommand{\N}[1]{\mathcal{N}\big(#1\big)}  
\newcommand{\GP}[1]{\mathcal{GP}\big(#1\big)}
\newcommand{\saemQ}{\mathbb{Q}}
\newcommand{\saemQf}{\saemQ^{\text{Fish}}}
\newcommand{\saemQb}{\saemQ^{\text{Bay}}}
\newcommand{\ei}{k}

\newcommand{\mi}{j}
\newcommand{\mI}{J}
\newcommand{\saS}{\mathbb{S}}

\newcommand{\kernel}[1]{\Pi_{#1}}
\newcommand{\ttheta}{{\widetilde\theta}}

\newcommand{\tX}{\widetilde x_{0:\T}} 
\newcommand\tx{\widetilde x}
\newcommand\ta{\widetilde a}
\newcommand\tw{\widetilde w}

\newcommand\tJ{{\widetilde J}}
\newcommand{\M}[2]{\mathsf{M}\!\left[#1, #2\right]}

\newcommand{\psaem}{PSAEM\xspace}

\newtheorem{theorem}{Theorem}
\newtheorem{lemma}{Lemma}

\begin{document}
%
\title{Learning dynamical systems with\\particle stochastic approximation EM}
%
%
%

\author{Andreas~Lindholm~
        and~Fredrik~Lindsten
\thanks{Andreas~Lindholm is with the Department	of Information Technology, Uppsala University, Sweden. E-mail: andreas.lindholm@it.uu.se.}
\thanks{Fredrik~Lindsten is with the Division of Statistics and Machine Learning, Linköping University, Sweden. E-mail: fredrik.lindsten@liu.se.}}

%
%

\markboth{Lindholm and Lindsten 2019: Learning dynamical systems with particle stochastic approximation EM}%
{Lindholm and Lindsten 2019: Learning dynamical systems with particle stochastic approximation EM}
%



\maketitle

\begin{abstract}
We present the particle stochastic approximation EM (PSAEM) algorithm for learning of dynamical systems. The method builds on the EM algorithm, an iterative procedure for maximum likelihood inference in latent variable models. By combining stochastic approximation EM and particle Gibbs with ancestor sampling (PGAS), PSAEM obtains superior computational performance and convergence properties compared to plain
particle-smoothing-based approximations of the EM algorithm. PSAEM can be used for plain maximum likelihood inference as well as for empirical Bayes learning of hyperparameters. Specifically, the latter point means that existing PGAS implementations easily can be extended with \psaem to estimate hyperparameters at almost no extra computational cost. We discuss the convergence properties of the algorithm, and demonstrate it on several signal processing applications.
\end{abstract}


%
\IEEEpeerreviewmaketitle

\section{Introduction}
%
%
%
%
\IEEEPARstart{L}{earning}  of dynamical systems, or state-space models, is central to many classical signal processing problems, such as
time-series modeling, filtering and control design.
State-space models are also at the core of recent model developments in machine learning, such as  
Gaussian process state-space models \cite{FrigolaCR:2014,MattosDD:2016}, infinite factorial dynamical models \cite{GaelTG:2009,ValeraRSP:2015}, and stochastic recurrent neural networks \cite[for example]{FraccaroSPW:2016}.
A strategy to learn state-space models, independently suggested by \cite{Digalakis1993} and \cite{GhahramaniH:1996}, is the use of the Expectation Maximization (EM) \cite{DempsterLR:1977} method. Originally proposed for maximum likelihood estimation of linear models with Gaussian noise, the strategy can be generalized to the more challenging non-linear and non-Gaussian cases, as well as the empirical Bayes setting. Many contributions have been made during the last decade, and this paper takes another step along the path towards a more computationally efficient method with a solid theoretical ground for learning of nonlinear dynamical systems.

To set the notation for this article, we write a general (discrete-time, non-linear and non-Gaussian) state-space model, or dynamical system, as
\begin{subequations}\label{eq:ssm}
	\begin{align}
	x_t &\sim p_\theta(x_t\mid x_{t-1}),\\
	y_t &\sim p_\theta(y_t\mid x_t),
	\end{align}
\end{subequations}
with transition density function $p_\theta(x_t \mid x_{t-1})$ and observation density function $p_\theta(y_t \mid x_t)$, parameterized by some unknown parameter $\theta \in \setTheta\subset\mathbb{R}^p$. Here, $\{ x_t \in \setX\subset\mathbb{R}^m\}_{t = 0, 1, \dots}$ denotes the unobserved state and $\{ y_t \in \setY\subset\mathbb{R}^d\}_{t = 1, 2 \dots}$ denotes the observations, and the index $t$ is referred to as 'time'.
The initial state $x_0$ is distributed according to\footnote{For notational brevity we assume that the initial density $p(x_0)$ is fully specified and not parameterized by $\theta$, but the extension to an unknown initial density is straightforward.} $p(x_0)$. We consider $\theta$ as unknown and the focus of this paper is to \emph{learn} it from recorded data $(y_1, ..., y_T) \triangleq \Y$.

The EM algorithm, which is the strategy we will follow, iteratively solves an integration problem and a maximization problem. When using EM for learning state-space models~\eqref{eq:ssm}, the integral includes the posterior distribution of the unobserved states $\X$, and possibly also the parameter $\theta$. This distribution is in general analytically intractable, but can be approximated using computational methods such as particle filters/sequential Monte Carlo (SMC).

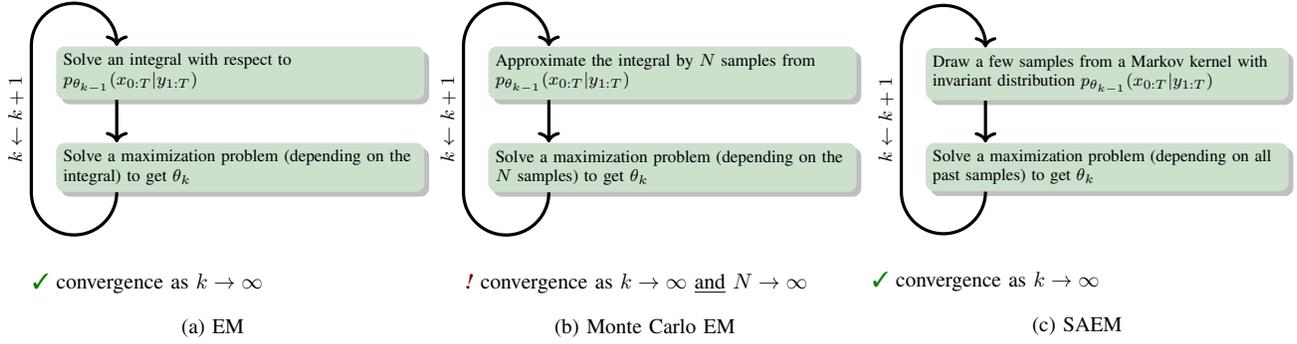
\begin{figure*}
	\centering
	\begin{subfigure}{.31\linewidth}
		\resizebox{\linewidth}{!}{%
			\begin{tikzpicture}
			\node[block,anchor=north west] (E) at (1,2.4) {Solve an integral with respect to $p_{\theta_{\ei-1}}(\X|\Y)$};
			\node[block,anchor=south west] (M) at (1,0) {Solve a maximization problem (depending on the integral) to get $\theta_{\ei}$};
			\node[anchor=south] (k) at (.6,1.2) [rotate=90] {\small $k\leftarrow k+1$};
			
			\draw[->,line width=1.5pt] ($(E.south west)+(1,0)$) to ($(M.north west)+(1,0)$);
			\draw[->,line width=1.5pt] let \p{A}=(k.south), \p{B}=($(M.south west)+(1,0)$), \p{C}=($(E.north west)+(1,0)$) in ($(M.south west)+(1,0)$) to [bend left=45] ($(M.south west)+(.3,-.7)$) to [bend left=45] (\x{A},\y{B}) to (\x{A},\y{C}) to [bend left=45] ($(E.north west)+(.3,.7)$) to [bend left=45] ($(E.north west)+(1,0)$);
			
			\node[anchor=west] (k) at (0.5,-1.5) {\textcolor{green!50!black}{\cmark} convergence as $k\to\infty$};
			\end{tikzpicture}}%
		\caption{EM}
	\end{subfigure}
	\begin{subfigure}{.31\linewidth}
		\resizebox{\linewidth}{!}{%
			\begin{tikzpicture}
			\node[block,anchor=north west] (E) at (1,2.4) {Approximate the integral by $N$ samples from $p_{\theta_{\ei-1}}(\X|\Y)$};
			\node[block,anchor=south west] (M) at (1,0) {Solve a maximization problem (depending on the $N$ samples) to get $\theta_{\ei}$};
			\node[anchor=south] (k) at (.6,1.2) [rotate=90] {\small $k\leftarrow k+1$};
			
			\draw[->,line width=1.5pt] ($(E.south west)+(1,0)$) to ($(M.north west)+(1,0)$);
			\draw[->,line width=1.5pt] let \p{A}=(k.south), \p{B}=($(M.south west)+(1,0)$), \p{C}=($(E.north west)+(1,0)$) in ($(M.south west)+(1,0)$) to [bend left=45] ($(M.south west)+(.3,-.7)$) to [bend left=45] (\x{A},\y{B}) to (\x{A},\y{C}) to [bend left=45] ($(E.north west)+(.3,.7)$) to [bend left=45] ($(E.north west)+(1,0)$);
			
			\node[anchor=west] (k) at (0.5,-1.5) [align=left] {\textcolor{red!50!black}{\textbf{\textit{!}}} convergence as $k\to\infty$ \underline{and} $N\to\infty$};
			\end{tikzpicture}}%
		\caption{Monte Carlo EM}
	\end{subfigure}
	\begin{subfigure}{.31\linewidth}
		\resizebox{\linewidth}{!}{%
			\begin{tikzpicture}
			\node[block,anchor=north west] (E) at (1,2.4) {Draw a few samples from a Markov kernel with invariant distribution $p_{\theta_{\ei-1}}(\X|\Y)$};
			\node[block,anchor=south west] (M) at (1,0) {Solve a maximization problem (depending on all past samples) to get $\theta_{\ei}$};
			\node[anchor=south] (k) at (.6,1.2) [rotate=90] {\small $k\leftarrow k+1$};
			
			\draw[->,line width=1.5pt] ($(E.south west)+(1,0)$) to ($(M.north west)+(1,0)$);
			\draw[->,line width=1.5pt] let \p{A}=(k.south), \p{B}=($(M.south west)+(1,0)$), \p{C}=($(E.north west)+(1,0)$) in ($(M.south west)+(1,0)$) to [bend left=45] ($(M.south west)+(.3,-.7)$) to [bend left=45] (\x{A},\y{B}) to (\x{A},\y{C}) to [bend left=45] ($(E.north west)+(.3,.7)$) to [bend left=45] ($(E.north west)+(1,0)$);
			
			\node[anchor=west] (k) at (0,-1.5) {\textcolor{green!50!black}{\cmark} convergence as $k\to\infty$};
			\end{tikzpicture}}%
		\caption{SAEM}
	\end{subfigure}
	\caption{A brief summary of the different flavors of Expectation Maximization (EM) methods for maximum likelihood estimation of $\theta$. When it is possible to solve the state inference problem (computing $p_{\theta_k}(\X|\Y)$) analytically, vanilla EM (a) is the preferred solution. For general models of the form~\eqref{eq:ssm}) this is not possible and numerical approximations are required. Monte Carlo EM (b) can be used with a particle smoother (an extension of the particle filter/SMC) to generate approximate samples from $p_{\theta_k}(\X|\Y)$, which has been proposed independently in the literature several times. Alternatively, stochastic approximation EM (SAEM, (c)) can be used (whose convergence properties are superior to Monte Carlo EM), and if the required Markov kernel is built up using the particle filter in \cref{alg:cpfas} we refer to it as \psaem, the main contribution of this paper.}
	\label{fig:intro}
\end{figure*}

The combination of EM and SMC, as suggested by \cite{CappeMR:2005,OlssonDCM:2008,SchonWN:2011}, has provided a principled solution to the challenging problem of learning general state-space models~\eqref{eq:ssm}, but is unfortunately `doubly asymptotic'; to ensure convergence, it requires (i) for \emph{each iteration} of the EM algorithm an infinite number of particles/Monte Carlo samples for approximating the posterior of $x_{0:T}$, and (ii) the EM algorithm itself converges only as its number of iterations goes to infinity. This is a theoretical as well as a practical issue, and we will in this paper explore a solution where particle Markov chain Monte Carlo (PMCMC), rather than plain SMC, is used, which allows the two asymptotical convergences to be `entangled'. This will give us an algorithm which relies on asymptotics only in one dimension (its number of iterations, not the number of particles), and thereby enjoys a significantly reduced computational cost and superior convergence properties compared to the predecessors. This overall picture is also briefly summarized in \cref{fig:intro}. 

Throughout the paper we assume that the reader is familiar with Markov chain Monte Carlo (MCMC, \cite{RobertC:2004,Tierney:1994}) as well as particle filters/SMC \cite{DoucetJ:2011,DoucetFG:2001}.

\section{Problem formulation and conceptual solution}\label{sec:problem}

Given a batch of observations $y_{1:\T}$ we wish to learn the unknown parameters $\theta$ as well as the unobserved states $x_{0:T}$ of the model~\eqref{eq:ssm}. For the states $\X$ we are interested in their posterior distribution.
For the parameters $\theta$, we 
consider two cases: In the frequentistic, or rather Fisherian, setting we are interested in a (possibly regularized) maximum likelihood estimate $\thml$. In the Bayesian setting, we assign a prior distribution to the parameters, $\theta \sim p_\eta(\theta)$. The prior, in turn, is assumed to be parameterized by some \emph{hyperparameter} $\eta$, which needs to be estimated\footnote{Hyperparameters can also be set using prior knowledge.}. Thus, we address both of the following two problems:
\begin{enumerate}\itemsep1em 
	\item \textbf{(Fisherian setting)} Compute the \emph{maximum likelihood estimate} of the model parameters,
	\[
	\thml = \argmax_\theta \log p_\theta(\Y),
	\]
	where \(p_\theta(\Y) = \int p_\theta(\Y\mid \X)p_\theta(\X)\myd\X\) is the likelihood function. A regularization term, such as $\|\theta\|_1$, may also be included in the maximization criterion.\footnote{To avoid too complicated expressions, we do not include the regularization term. From a user's perspective, it simply amounts to replace $p_\theta(\Y\mid \X)$, whenever it appears, with $p_\theta(\Y\mid \X)+\lambda \frac{1}{n}\|\theta\|_1$.}
	\item \textbf{(Bayesian setting)} Compute the \emph{posterior distribution} of the model parameters $p_\etaml(\theta\mid \Y)$, where the hyperparameters are estimated using empirical Bayes
	\[
	\etaml = \argmax_\eta \log p_\eta(\Y),
	\]
	where the \emph{marginal} likelihood function is \(p_\eta(\Y) = \int p_\theta(\Y\mid \X)p_\theta(\X)p_\eta(\theta)\myd\theta\myd\X\).	
\end{enumerate}
These two problems are in fact strongly related, and can be seen as exactly the same problem on a more abstract level. In the interest of concreteness we will, however, distinguish between those two problems, but the computational algorithm that we will propose can be used to address both. The algorithm will compute a Monte Carlo approximation of the posterior distribution over the latent variables. 

Both settings involve the computation of a maximum likelihood estimate; of the model parameters in the first case and of the hyperparameters in the second case. A conceptual solution to these problems is given by the expectation maximization (EM, \cite{DempsterLR:1977}) algorithm. EM is a data augmentation method, meaning that it is based on the notion of a \emph{complete data}, comprising the observed data as well as the latent (or missing) variables. The EM algorithm iteratively updates the \mbox{(hyper-)}parameters, and each iteration consists of two steps:
\begin{itemize}
	\item[(E)] Compute the expected value of the complete data log-likelihood for fixed \mbox{(hyper-)}parameters
	\item[(M)] Maximize the $Q$-function (which will be defined below) with respect to the (hyper-)parameters
\end{itemize}
The observed data is always $\Y$, and what differs between the Fisherian and the Bayesian problem is what constitutes the latent variables. For the Fisherian problem, the latent variables are $\X$ and we obtain, at iteration $\ei$,

\vspace{-1em}

\small
	\begin{align}\small
	\text{(E)}&& &\text{Let }\Qf_k(\theta) \eqdef \int \log p_\theta(\Y, \X) p_{\theta_{\ei-1}}(\X \mid \Y) \myd \X,\nonumber\\
	\text{(M)}&& &\text{Solve }\theta_{\ei} \leftarrow \argmax_{\theta} \Qf_k(\theta).\label{eq:EMf}
	\end{align}

\normalsize
Note that the expectation in the (E)-step is \wrt the smoothing distribution $p_{\theta_{\ei-1}}(\X \mid \Y)$ parameterized by the previous parameter iterate $\theta_{\ei-1}$. It is well known  that iterating~\eqref{eq:EMf} gives a monotone increase of the likelihood $p_{\theta_k}(\Y)$, and $\theta_\ei$ will under weak assumptions converge to a stationary point of the likelihood function as $\ei\to\infty$ (\eg \cite{Wu:1983}).

For empirical Bayes we obtain similar expressions, but the latent variables are \emph{both} $\X$ \emph{and} $\theta$. The (E)-step is as

\vspace{-1.3em}

\small
	\begin{align}\nonumber
	\text{(E)}~&\text{Let }\Qb_k(\eta) \nonumber \eqdef \\ & \int \log p_\eta(\Y, \X, \theta) p_{\eta_{\ei-1}}(\theta, \X \mid \Y) \myd \theta \myd \X = \nonumber\\ 
	&\int \log p_\eta(\theta) p_{\eta_{\ei-1}}(\theta \mid \Y) \myd \theta + \text{const.},\nonumber\\
	\text{(M)}~&\text{Solve }\eta_{\ei} \leftarrow \argmax_{\eta} \Qb_k(\eta),\label{eq:EMb}
	\end{align}

\normalsize\noindent 
where the second line of~\eqref{eq:EMb} follows from the fact that in the factorization of the complete data likelihood, only the prior density $p_\eta(\theta)$ depends on the hyperparameter $\eta$.
The M-step remains unchanged. In complete analogy to the Fisherian setting,~\eqref{eq:EMb} will also under weak assumptions converge to a stationary point of the marginal likelihood as $\ei\to\infty$.

Both~\eqref{eq:EMf} and~\eqref{eq:EMb} can be implemented and iterated until convergence, as long as the integrals can be computed and the maximization problem solved. However, in most cases---specifically for the models we consider in this paper---the integrals can \emph{not} be solved analytically, and the topic for the rest of this paper is essentially to design an efficient method for approximating the integrals. The solution will be based on PMCMC \cite{AndrieuDH:2010}, but also a stochastic approximation of the $Q$-function \cite{DelyonLM:1999} to ensure a computationally efficient solution with good convergence properties. Our solution will therefore be more involved than just replacing the integrals in~\eqref{eq:EMf} or~\eqref{eq:EMb} with vanilla Monte Carlo estimators (Monte Carlo EM).

A short word on notation: we will use subscripts to denote sequences of variables for which we are seeking a maximum, like $\eta_\ei$, and brackets for samples of variables for which we are seeking a posterior distributions, like $\X[\ei]$.

\section{Related work and contributions}\label{sec:rw}%
The use of EM for learning linear state-space models appears to have been independently suggested by, at least, \cite{Digalakis1993} and \cite{GhahramaniH:1996}. For linear models the state inference problem can be solved exactly using a Kalman filter, but not for nonlinear models. To this end, the extended Kalman filter has been proposed \cite{GR:1999,DL:2013}, as well as SMC-based solutions \cite{CappeMR:2005,OlssonDCM:2008,SchonWN:2011}, leading to a so-called Monte Carlo EM solution. 

EM is a general strategy for latent variable models, and the standard choice in the application of EM to state-space models is to select the states as the latent variable. It is, however, shown by \cite{UmenbergerWM:2018} that if considering the process noise realization (instead of the states) as latent variables, it is possible to introduce stability guarantees for the learned model at the cost of a more involved maximization problem.

This paper considers the offline (or batch) problem, but EM can also be applied for online (or streaming data) problems. For the nonlinear online problem, the combination of EM and SMC dates back to at least \cite{AndrieuD:2003,AndrieuDT:2005}, and recent contributions include \cite{OW:2015}.

Stochastic approximation EM (SAEM, \cite{DelyonLM:1999,KuhnL:2004}) can be used to improve the convergence properties and reduce the computational cost, compared to Monte Carlo EM. 
This is particularly true when the Monte Carlo simulation is computationally involved, which is the case for SMC-based solutions.
In the context of state-space models, SAEM appears to first have been proposed by \cite{DonnetS:2011} and \cite{AndrieuV:2014}, who suggest to combine it with a particle independent Metropolis--Hastings procedure (PIMH, \cite{AndrieuDH:2010}) to infer the latent states. The idea to combine SAEM with particle Gibbs with ancestor sampling (PGAS, \cite{LindstenJS:2014}), which often has a much lower computational cost, was first suggested in a brief conference paper by \cite{Lindsten:2013}---the present article is an extension of this paper. Since its first publication, this method---which we refer to as \psaem--- has found applications in system identification \cite{SS:2017,SvenssonLS:2014}, causal inference \cite{GongZS:2017}, and econometrics \cite{SingorBA:2017}, to mention a few. In this paper, we will study \psaem more thoroughly, formulate it explicitly for empirical Bayes, 
present a new theoretical result, and illustrate the method's applica\-bility to some contemporary dynamical systems models from the machine learning literature (Gaussian process state-space models \cite{FrigolaCR:2014,SSS+:2016} and infinite factorial dynamical models \cite{ValeraRSP:2015}).

\section{Particle stochastic approximation EM}\label{sec:psaem}
We will now build up and present the contribution of this paper, the particle stochastic approximation EM (\psaem) algorithm. The two main components are (i) an MCMC kernel for simulating the latent variables from either $p_\theta(\X\mid \Y)$ or $p_\eta(\theta, \X \mid \Y)$, and (ii) a stochastic approximation version of the EM algorithm (SAEM, \cite{DelyonLM:1999,KuhnL:2004}), to update the (hyper-)parameter estimate. We will start with the former (\cref{sec:PGAS}) and thereafter turn to the latter (\cref{sec:SAEM}).

\subsection{Sampling the latent variables using PGAS}\label{sec:PGAS}
At the core of the EM algorithm is the posterior inference of the latent variables, which is needed for the integrals in \eqref{eq:EMf} or \eqref{eq:EMb}. For general non-linear or non-Gaussian state-space models these posterior distributions are intractable and we have to use numerical approximations. Much research has been done over the past decades on computational algorithms for this problem, and many powerful tools are available. We will focus on PMCMC \cite{AndrieuDH:2010} methods which we believe are particularly well suited for this. PMCMC is a framework for using particle filters to construct efficient high-dimensional Markov kernels, and we will specifically make use of the particle Gibbs with ancestor sampling (PGAS) \cite{LindstenJS:2014}. PGAS has been shown to have good empirical performance in many situations (e.g., \cite{LindermanSA:2014,MeentHMW:2015,ValeraRSP:2015,MarcosCBD:2015}), but other versions of particle Gibbs could possibly also be employed, such as Particle Gibbs with backward simulation \cite{Whiteley:2010,LindstenS:2012} or blocked particle Gibbs \cite{SinghLM:2015}.

To start we assume that the parameters to estimate ($\theta$ or $\eta$) are fixed at some value, and consider how $p_\theta(\X\mid \Y)$ or $p_\eta(\theta,\X\mid\Y)$ can be approximated using PGAS. Consider first the Fisherian setting. Just like any MCMC method would do, PGAS makes use of an Markov kernel on the space $\setX^{\T+1}$ with $p_\theta(\X\mid \Y)$ as its unique stationary distribution. This kernel is then applied iteratively, and if certain ergodicity assumptions hold, this procedure will eventually produce samples from $p_\theta(\X\mid \Y)$. With PGAS this Markov kernel is constructed using a particle filter, or more precisely a \emph{conditional particle filter with ancestor sampling}, given in \cref{alg:cpfas}. One execution of the entire \cref{alg:cpfas} corresponds to one iteration of the Markov kernel. The conditional particle filter resembles a standard particle filter with $N-1$ particles, with the addition that there is also a \emph{conditional} particle trajectory (for convenience numbered $N$, line 2 and 8) which is specified a priori. In the resampling step (line 5), this conditional trajectory can be replicated, but never discarded. At the end, one single trajectory is extracted, which will be used as conditional trajectory in a later iteration. The ancestor sampling (line~7) assigns ancestors to the conditional trajectory, similar to resampling but `backwards' in time and only for the conditional trajectory. We refer to \cite{LindstenJS:2014} for further details. \Cref{alg:cpfas} is formulated in its `bootstrap' version, but a more general SMC formulation is also possible, see \cite{LindstenJS:2014}.

\begin{algorithm}\algfontsize
	\begin{algorithmic}[1]
		\Statex \textbf{Input:} Conditional trajectory $\X'$, parameter $\theta$.
		\Statex \textbf{Output:} Trajectory $\X^\star$.
		\State Draw $x_0^i \sim p(x_0), i = 1, \dots, N-1$.
		\State Set $x_0^N \leftarrow x'_0$.
		\State Set $w_0^i \leftarrow 1, i = 1, \dots, N$.
		\For{$t = 1, 2, \dots, T$}
		\State Draw $a_t^i$ with $\Pr(a_t^i = j) \propto w_{t-1}^j$ for $i = 1, \dots, N-1$.
		\State Draw $x_t^i\sim p_\theta(x_t\mid x_{t-1}^{a_t^i})$ for $i = 1, \dots, N-1$.
		\State Draw $a_t^N$ with $\Pr(a_t^N = j) \propto w_{t-1}^jp_\theta(x'_t\mid x_{t-1}^j)$.
		\State Set $x_t^N \leftarrow x_t'$.
		\State Set $w_t^i \leftarrow p_\theta(y_t\mid x_t^i)$ for $i = 1, \dots, N$.
		\EndFor
		\State Draw $I$ with $\Pr(I = i) \propto w_{T}^i$.
		\State Set $x^\star_T = x^I_T$.
		\For{$t=T-1, T-2, \dots, 0$}
		\State Set $I \leftarrow a^I_{t+1}$.
		\State Set $x^\star_t \leftarrow x^{I}_t$.
		\EndFor
	\end{algorithmic}
	\caption{Markov kernel $\kernel{\theta}(\X',\X^\star)$}
	\label{alg:cpfas}
\end{algorithm}

Formally we let \cref{alg:cpfas} define a Markov kernel $\kernel{\theta}$ on the space of state trajectories $\setX^{T+1}$ given by
\begin{align}
\label{eq:Pi-X-def}
\kernel{\theta}(\X^\prime, B) =  \E\left[\I(\X^\star \in B )\right] 
\end{align}
where the expectation is \wrt the random variables used in Algorithm~\ref{alg:cpfas}. 
The Markov kernel constructed by \cref{alg:cpfas} takes a state trajectory $\X[\mi-1] = \X' \in \setX^{\T+1}$ as input and outputs another state trajectory $\X[\mi] = \X^\star \in \setX^{\T+1}$.
Put differently, a sample $\X[\mi] \sim \kernel{\theta}(\X[\mi-1], \cdot)$ can be generated by executing \cref{alg:cpfas} with fixed $\theta$ and $\X[\mi-1]$ as input reference trajectory. If this is iterated, an MCMC procedure on the space $\setX^{T+1}$ is obtained, and the trajectories $\X[0]$, $\X[1]$, $\X[2]$, \dots, will eventually be samples from the sought smoothing distribution $p_\theta(\X\mid \Y)$. MCMC methods like this, which uses Markov kernels based on particle filters, are called PMCMC.

It is far from obvious that $\kernel{\theta}$ 
admits $p_\theta(\X\mid \Y)$ as its stationary distribution. However, its properties (as well as those of its older sibling presented by \cite{AndrieuDH:2010}) have been extensively studied, see for example \cite{AndrieuDH:2010,ChopinS:2015,LindstenJS:2014,LindstenDM:2015,AndrieuLV:2018,DelMoralKP:2014}. The main results are: (i) $p_\theta(\X\mid \Y)$ is a stationary distribution of $\kernel{\theta}$, and
(ii) $\kernel{\theta}$ is uniformly geometrically ergodic in $\X$ for any $N\geq 2$ and any $\theta\in\Theta$ under upper boundedness conditions on $w_t^i$ in \cref{alg:cpfas}. We summarize this PMCMC procedure to infer $p_\theta(\X\mid \Y)$ in \cref{alg:mcmcf} (still assuming $\theta$ is fixed).
\vspace{-.7em}
\begin{algorithm}\algfontsize
	\caption{Sampling $p_\theta(\X\mid \Y)$ (Fisherian, fixed $\theta$)}
	\begin{algorithmic}[1]
		\State Initialize $\X[0]$ arbitrarily, e.g., by running a standard particle filter targeting $p_\theta(\X\mid \Y)$.
		\For{$\mi = 1, 2, \dots, \mI$}
		\State Sample $\X[\mi]\sim \kernel{\theta}( \X[\mi-1], \cdot)$ (run Alg.~\ref{alg:cpfas} once)
		\EndFor
	\end{algorithmic}
	\label{alg:mcmcf}
	
\end{algorithm}
\vspace{-.7em}

So far we have only considered the Fisherian setting, in which the latent variables only comprise the state trajectory. For the Bayesian setting we assume that $\eta$ (instead of $\theta$) is fixed, and we see from \eqref{eq:EMb} that we have to compute the model parameter posterior distribution $p_\eta(\theta\mid\Y)$. We will do this by splitting the simulation problem into two steps, one in which we sample $\X$ conditionally on $\theta$ (and $\Y$) and one in which we sample $\theta$ conditionally on $\X$ (and $\Y$). The first step, sampling $\X$ conditionally on $\theta$, is equivalent to the problem discussed for the Fisherian setting, and we can use the Markov kernel $\kernel{\theta}$ \cref{alg:cpfas}. For the second step, sampling $\theta$ conditionally on $\X$, exact solutions are often possible, leading to Gibbs sampling. Otherwise, methods like Hastings-within-Gibbs (see, for instance, \cite[Section 2.4]{Tierney:1994}) are possible. The particular choice depends on the actual model, and we will later illustrate it by an example. Let the Markov kernel used to simulate $\theta$ be denoted by $\Pi_{\eta,\X}(\theta', \cdot)$. Most of the previously referenced literature on properties for $\kernel{\theta}$ covers also the setting of a joint kernel for $\theta,\X$.
The resulting MCMC procedure used in the Bayesian setting (still assuming a fixed value for $\eta$) is summarized in \cref{alg:mcmcb}, and converges (in the same sense as \cref{alg:mcmcf}) to $p_\eta(\theta,\X\mid \Y)$.

\begin{algorithm}\algfontsize
	\caption{Sampling $p_\eta(\theta,\X\mid \Y)$ (Bayesian, fixed $\eta$)}
	\begin{algorithmic}[1]
		\State Initialize $\theta[0]$ and $\X[0]$ arbitrarily. For the latter e.g., by a particle filter targeting $p_{\theta[0]}(\X\mid \Y)$
		\For{$\mi = 1, 2, \dots$}
		\State Sample $\X[\mi]\sim \kernel{\theta[\mi-1]}( \X[\mi-1], \cdot)$ (run Alg.~\ref{alg:cpfas} once)
		\State Sample $\theta[\mi]\sim \Pi_{\eta,\X[\mi]}(\theta[\mi-1], \cdot)$
		\EndFor
	\end{algorithmic}
	\label{alg:mcmcb}
\end{algorithm}
\vspace{-.7em}

\subsection{Combining PGAS and EM}\label{sec:SAEM}

We have so far assumed that the (hyper-)parameters are fix. The objective in this paper is, however, to learn those, and we will for this purpose use a stochastic approximation version of the EM algorithm.

\subsubsection{A naive solution using EM and PMCMC}

The problem with the preliminary EM solutions outlined in~\eqref{eq:EMf} and~\eqref{eq:EMb}, respectively, is the analytically intractable integrals in their $Q$-functions. A first idea would be to replace 
the integrals with sums over $\mI$ Monte Carlo samples. For the Fisherian setting, this means replacing the (E)-step of~\eqref{eq:EMf} with a simulation (Si) step as follows:

\vspace{-1em}

\small
	\begin{align}
	\text{(Si)}&& &\begin{cases}
	\text{Draw }&\{\X[\mi]\}_{\mi=1}^\mI \sim ~p_{\theta_{\ei-1}}(\X \mid \Y)  \\
	\text{and let }&\Qaf_k(\theta) \eqdef \frac{1}{\mI}\sum_{\mi=1}^{\mI} \log p_\theta(\Y, \X[\mi]).\end{cases} \nonumber \\
	\text{(M)}&& &\text{Solve }\theta_{\ei} \leftarrow \argmax_{\theta} \Qaf_k(\theta).\label{eq:MCEMf}
	\end{align}

\normalsize
Note that this is our initial EM scheme~\eqref{eq:EMf}, but with the analytically intractable integral over $\log p_\theta(\Y, \X)$ approximated by a sum. This algorithm is commonly referred to as Monte Carlo EM \cite{WeiT:1990} or, if $\mI=1$, stochastic EM \cite{DieboltI:1996}. To draw the samples in the (Si)-step we can use PGAS from \cref{sec:PGAS}, which would give \cref{alg:naiveEM}.
\vspace{-.7em}
\begin{algorithm}[H]\algfontsize
	\caption{Monte Carlo EM for the Fisherian problem}
	\begin{algorithmic}[1]
		\State Initialize $\theta_0$
		\For{$\ei = 1, 2, \dots$}
		\State\label{alg:naiveEM:E} Run Alg.~\ref{alg:mcmcf} with $\theta_{\ei-1}$ `until convergence', get $\{\X[\mi]\}_{\mi=1}^\mI$
		\State\label{alg:naiveEM:M} Solve $\theta_\ei \leftarrow \argmax_{\theta}  \frac{1}{\mI} \sum_{\mi=1}^{\mI} \log p_\theta(\Y, \X[\mi])$
		\EndFor
	\end{algorithmic}
	\label{alg:naiveEM}
\end{algorithm}

\vspace{-.7em}

\noindent A similar algorithm could be devised for the Bayesian setting. Even though \cref{alg:naiveEM} might look promising, there are two issues with this solution:
\begin{enumerate}[(i)]\setlength{\itemsep}{0em}
	\item To guarantee that \cref{alg:mcmcf} has converged to its stationary distribution, we cannot bound its number of iterations at line 3.
	\item For the sum in~\eqref{eq:MCEMf}/line 4 to converge to the integral it approximates, we must let $\mI\to\infty$.
\end{enumerate}
Indeed, these two issues are related. We basically need to allow $\mI\to\infty$ to ensure convergence, whilst the convergence of the EM iteration happens as $\ei\to\infty$. This is not desirable since it, intuitively, gives a computational complexity of ``$\infty\times\infty$''; see further \cite{FortM:2003}. 
Existing methods based on various types of particle smoothing for approximating the integral with respect to $p_\theta(\X\mid\Y)$, for instance \cite{OlssonDCM:2008,SchonWN:2011}, suffer from the same issues. Indeed, these methods are (SMC-based) instances of Monte Carlo EM.

We will now first address issue (ii) with stochastic approximation EM, and thereafter handle issue (i) by `entangling' the convergence of \cref{alg:mcmcf} ($\mI\to\infty$) with the convergence of the EM algorithm ($\ei\to\infty$).

\subsubsection{SAEM: Handling sample approximations within EM}

Stochastic approximation, as introduced by \cite{RM:1951}, is an averaging procedure to solve a (deterministic) equation which can only be evaluated through noisy (stochastic) observations. In stochastic approximation a step length $\gamma_k\in[0,1]$ is used, 
which has to fulfill 
\begin{align}
\sum_{k=1}^{\infty}\gamma_k = \infty, \qquad \sum_{k=1}^{\infty}\gamma_k^2 < \infty, \qquad \gamma_1 = 1.\label{eq:gamma}
\end{align}
Following \cite{DelyonLM:1999}, the SAEM algorithm can be introduced by making a stochastic approximation of the $Q$-function.
In SAEM, we transform Monte Carlo EM~\eqref{eq:MCEMf} by introducing a stochastic approximation (SA)-step. For simplicity we only use one sample ($\mI=1$) in the simulation (Si)-step, but in practice it can be favorable to use a small batch of samples. For iteration $\ei$ this becomes

\vspace{-1em}

\small
	\begin{align}
	\text{(Si)}&&& \text{Draw }\X[\ei] \sim ~p_{\theta_{\ei-1}}(\X \mid \Y).\nonumber\\
	\text{(SA)}&&& \text{Let }\saemQf_k(\theta) \leftarrow (1-\gamma_k)\saemQf_{\ei-1}(\theta) + \gamma_k \log p_\theta(\Y, \X[\ei]).\nonumber\\
	\text{(M)}&& &\text{Solve }\theta_{\ei} \leftarrow \argmax_{\theta} \saemQf_k(\theta).\label{eq:SAEMf}
	\end{align}

\normalsize
To intuitively understand the stochastic approximation, let us first ignore the (M)-step and assume $\gamma_\ei = \frac{1}{\ei}$. In such a case, the (SA)-step would simply be online averaging, equivalent to $\saemQf_\ei(\theta) = \frac{1}{\ei}\sum_{\ell=1}^{\ei} \log p_\theta(\Y, \X[\ell])$, where $\X[\ell]\sim p_{\theta}(\X \mid \Y)$, which 
converges to $\int \log p_\theta(\Y, \X)p_{\theta}(\X \mid \Y)\myd\X$ when $\ei\to\infty$ by the law of large numbers. The introduction of the (M) step complicates the picture, but assuming that $\theta_\ei$ will eventually converge to a stationary point, the influence from the transient phase will vanish as $\ei\to\infty$, and the averaging argument can still be applied. In \cref{sec:conv} we discuss the convergence properties in detail. Before that, in \cref{sec:ef}, we will consider the important special case of exponential family models, for which the (SA) step reduces to a convenient recursive update of sufficient statistics.

With~\eqref{eq:SAEMf} in place, we can make stronger theoretical claims (even though we are using only a single sample, $\mI=1$, from $p_{\theta_{\ei-1}}(\X \mid \Y)$, at each iteration!) thanks to the use of stochastic approximation \cite{DelyonLM:1999}. However, for the problem under study it is still of limited practical use since we cannot generate samples from $p_{\theta_{\ei-1}}(\X \mid \Y)$ by other means than using~\cref{alg:mcmcf} with an infinite number of iterations (in order to ensure that it has converged). Thus, our final step is to use the method studied by \cite{KuhnL:2004} to combine SAEM with an MCMC procedure in a more intricate way than~\eqref{eq:SAEMf}.

\subsubsection{PSAEM: Combining SAEM with PGAS}

As suggested and analyzed by \cite{KuhnL:2004}, the draw from $p_{\theta_{\ei-1}}(\X \mid \Y)$ in~\eqref{eq:SAEMf} can be replaced with a draw from a Markov kernel which has $p_{\theta_{\ei-1}}(\X \mid \Y)$ as its invariant distribution. As discussed, this is exactly what $\kernel{\theta}$ from \cref{alg:cpfas} is, and we can thus assemble

\vspace{-1em}

\small
	\begin{align}
	\text{(Si)}&&& \text{Draw }\X[\ei] \sim ~\kernel{\theta_{\ei-1}}( \X[\ei-1], \cdot) \nonumber\\ &&&\text{ (that is, run Algorithm~\ref{alg:cpfas} \emph{once})}.\nonumber\\
	\text{(SA)}&&& \text{Let }\saemQf_k(\theta) \leftarrow (1-\gamma_k)\saemQf_{\ei-1}(\theta) + \gamma_k \log p_\theta(\Y, \X[\ei]).\nonumber\\
	\text{(M)}&& &\text{Solve }\theta_{\ei} \leftarrow \argmax_{\theta} \saemQf_k(\theta).\label{eq:PSAEMf}
	\end{align}

\normalsize
Note that we do not make use of \cref{alg:mcmcf} anymore, but only \cref{alg:cpfas}. This means that we do not run the Markov kernel ``until convergence'' at each iteration, but it will (intuitively speaking) converge in parallel with the SAEM iterations indexed with $\ei$. We summarize and present this as~\cref{alg:psaemf,alg:psaemb}.

\vspace{-.7em}

\begin{algorithm}[H]\algfontsize
	\caption{PSAEM for the Fisherian setting}
	\begin{algorithmic}[1]
		\State Initialize $\X[0], \theta_0$
		\For{$\ei = 1, 2, \dots$}
		\State Run Alg.~\ref{alg:cpfas} cond. on $\X[\ei-1]$ and $\theta_{\ei-1}$ to sample $\X[\ei]$
		\State $\saemQf_k(\theta) \leftarrow (1-\gamma_k)\saemQf_{\ei-1}(\theta) + \gamma_k \log p_\theta(\Y, \X[\ei])$
		\State Solve and update parameters $\theta_{\ei} \leftarrow \argmax_{\theta} \saemQf_k(\theta)$
		\EndFor
	\end{algorithmic}
	\label{alg:psaemf}
\end{algorithm}

\vspace{-.7em}

\begin{algorithm}[H]\algfontsize
	\caption{PSAEM for the Bayesian setting}
	\begin{algorithmic}[1]
		\State Initialize $\X[0], \theta[0], \eta_0$
		\For{$\ei = 1, 2, \dots$}
		\State Run Alg.~\ref{alg:cpfas} cond. on $\X[\ei$-$1]$ and $\theta[\ei$-$1]$ to sample $\X[\ei]$
		\State Sample $\theta[\ei]\sim \Pi_{\eta_{\ei-1}, \X[\ei]}(\theta[\ei-1], \cdot)$ 
		\State Update $\saemQb_k(\eta) \leftarrow (1-\gamma_k)\saemQb_{\ei-1}(\eta) + \gamma_k \log p_\eta(\theta[\ei])$
		\State Solve and update hyperparameters $\eta_{\ei} \leftarrow \argmax_{\eta} \saemQb_k(\eta)$
		\EndFor
	\end{algorithmic}
	\label{alg:psaemb}
\end{algorithm}

\vspace{-.7em}

We have now obtained an algorithm which only relies on asymptotics as $k\to\infty$, by `entangling' the convergence of PGAS with the convergence of SAEM. As we will see in \cref{sec:conv}, convergence can be shown under certain assumptions. We will now consider the important special case of models~\eqref{eq:ssm} in the exponential family, for which the recursively defined function $\saemQ_\ei$ reduces to a much simpler expression.

\subsection{\psaem for exponential family models}\label{sec:ef}

Studying \cref{alg:psaemf} or~\ref{alg:psaemb}, one may expect the computational cost of all computations involving the $\saemQ$-function to increase as $\ei\to\infty$, since $\saemQ_k$ is defined as a sum with $k$ terms, each a function of a past sample of $\X$. This is, however, not the case if the model belongs to the exponential family, which is an important special case discussed below.

When we write ``the model belongs to the exponential family'', we mean that the joint distribution for the latent and observed variables, $p_\theta(\X,\Y)$ or $p_\eta(\theta,\X,\Y)$, belongs to the exponential family with $\theta$ or $\eta$ as its parameter, respectively. For the Fisherian case, this is fulfilled if both equations in~\eqref{eq:ssm} can, with some choice of $S_x: \setX\times\setX\to\mathbb{R}^{\ell},\psi_x:\Theta\to\mathbb{R},\phi_x:\Theta\to\mathbb{R}^\ell$ (for some $\ell$), and similarly for some $S_y,\psi_y,\phi_y$, be written as

\vspace{-1em}

\small
\begin{subequations}\label{eq:expff}
	\begin{align}
	p_\theta(x_t\mid x_{t-1}) &\overset{\theta}{\propto} \exp\left\{-\psi_x(\theta)+\langle S_x(x_{t-1}, x_t),\phi_x(\theta)\rangle\right\},\\
	p_\theta(y_t\mid x_t) &\overset{\theta}{\propto} \exp\left\{-\psi_y(\theta)+\langle S_y(x_t, y_t),\phi_y(\theta)\rangle\right\}.
	\end{align}
\end{subequations}

\normalsize
Here, $\overset{\theta}{\propto}$ reads ``proportional (with respect to $\theta$) to'' and  $\langle\cdot,\cdot\rangle$ is an inner product. The subscripts ($x$, $y$ and $\theta$) do not denote dependencies in this context, but are only names. 

For the Bayesian case, the requirements are weaker, and it is enough that the prior distribution for $\theta$ belongs to the exponential family, 

\vspace{-1em}

\small
\begin{align}
p_\eta(\theta) &\overset{\eta}{\propto} \exp\left\{-\psi_\theta(\eta)+\langle S_\mathsf{\theta}(\theta),\phi_\theta(\eta)\rangle\right\}.\label{eq:expfb}
\end{align}

\normalsize
We will now see how the $\saemQ$-function from the (SA)-step simplifies for models which can be written on one of these forms.
First consider the Fisherian case. Using the Markovian structure of~\eqref{eq:ssm}, we can write

\vspace{-1em}

\small
\begin{align}
\log p_\theta(y_{1:T}, x_{0:T}) &= \sum_{t=1}^T \log p_\theta(y_t\mid x_t) + \log p_\theta(x_t\mid x_{t-1}) + \text{const.} \nonumber\\
&= -\psi(\theta)+\langle S(\X,\Y),\phi(\theta)\rangle + \text{const.}\label{eq:expf2}
\end{align}
{\normalsize where}
\begin{align*}
\psi(\theta) &= T\left\{\psi_x(\theta)+\psi_y(\theta)\right\},\\
S(\X,\Y) &= \sum_{t=1}^T \begin{pmatrix}
S_x(x_{t-1}, x_t) \\ S_y(x_t, y_t)
\end{pmatrix}, \\
\phi(\theta) &= \begin{pmatrix}
\phi_x(\theta) \\ \phi_y(\theta)
\end{pmatrix}.
\end{align*}

\normalsize
Here we have used the fact that the initial distribution $p(x_0)$ is independent of $\theta$ (for notational simplicity).
It follows that
\begin{subequations}
	\begin{align}
	\saemQf_k(\theta) = -\psi(\theta)+\langle \saS_k,\phi(\theta)\rangle  + \text{constant},\label{eq:saQ}
	\end{align}
	where
	\begin{align}
	\saS_k = (1-\gamma_k)\saS_{k-1} + \gamma_kS(\X[\ei],\Y).\label{eq:saS}
	\end{align}
\end{subequations}

Note that this is a non-recursive definition of $\saemQf_k(\theta)$, but instead recursive in $\saS_k$. From an algorithmic point of view, this means that we can compute and store $\saS_k$ as~\eqref{eq:saS}, and solve the maximization problem for~\eqref{eq:saQ} instead of the more intricate and computationally challenging~\eqref{eq:PSAEMf}. In fact, the maximizing argument to~\eqref{eq:saQ} can be expressed on closed form in many cases.

Analogously the Bayesian case is obtained as,
\begin{subequations}
	\begin{align}
	\saemQb_k(\eta) = -\psi_\theta(\eta)+\langle \saS_k,\phi_\theta(\eta)\rangle  + \text{constant}.\label{eq:saQb}
	\end{align}
	where
	\begin{align}
	\saS_k = (1-\gamma_k)\saS_{k-1} + \gamma_k S_\mathsf{\theta}(\theta[\ei]).\label{eq:saSb}
	\end{align}
\end{subequations}
We summarize in \cref{alg:psaemfexp,alg:psaembexp}.

\begin{algorithm}[H]\algfontsize
	\caption{PSAEM for exponential family models, Fisherian}
	\begin{algorithmic}[1]
		\State Initialize $\X[0]$ and $\theta_0$
		\For{$\ei = 1, 2, \dots$}
		\State Run Algorithm~\ref{alg:cpfas} conditional on $\X[\ei-1]$ and $\theta_{\ei-1}$ to sample $\X[\ei]$
		\State Update sufficient statistics $\saS_k$ according to \eqref{eq:saS}
		\State Solve and update parameters $\theta_{\ei} \leftarrow \argmax_{\theta}\saemQf_\ei(\theta)$ using \eqref{eq:saQ}
		\EndFor
	\end{algorithmic}
	\label{alg:psaemfexp}
\end{algorithm}

\vspace{-.7em}

\begin{algorithm}[H]\algfontsize
	\caption{PSAEM for exponential family models, Bayesian}
	\begin{algorithmic}[1]
		\State Initialize $\X[0]$, $\theta[0]$ and $\eta_0$
		\For{$\ei = 1, 2, \dots$}
		\State Run Algorithm~\ref{alg:cpfas} conditional on $\X[\ei-1]$ and $\theta[\ei-1]$ to sample $\X[\ei]$
		\State Sample $\theta[\ei]\sim \Pi_{\eta_{\ei-1}, \X[\ei]}(\theta[\ei-1], \cdot)$
		\State Update sufficient statistics $\saS_k$ according to \eqref{eq:saSb}
		\State Solve and update hyperparameters $\eta_{\ei} \leftarrow \argmax_{\eta}\saemQb_\ei(\eta)$ using \eqref{eq:saQb}
		\EndFor
	\end{algorithmic}
	\label{alg:psaembexp}
\end{algorithm}

\section{Convergence}\label{sec:conv}

The convergence of SAEM and its extensions, including MCMC-based implementations, has received a lot of attention \cite{DelyonLM:1999,KuhnL:2004,AndrieuMP:2005,AndrieuV:2014}. In \cref{sec:theory} we present a basic convergence result for \psaem. This is essentially an application of \cite[Theorem~1]{KuhnL:2004},
however, we also add a missing piece regarding the continuity of the PGAS Markov kernel. This will under certain (strong) assumptions on $\setX$ and the model~\eqref{eq:ssm} imply convergence of \psaem as $\ei\to\infty$ (with finite $N\geq2$ fixed in \cref{alg:cpfas}).
Some of these conditions could possibly be weakened by using the algorithmic modifications proposed by \cite{AndrieuV:2014}, but we do not pursue this further here.
We will also, in \cref{sec:practice}, discuss some practical considerations regarding the choice of $N$ and~$\gamma_k$.

In the presentation below we write $\|f\|_\infty = \sup_x|f(x)|$ for the supremum norm of function $f$ and $\Pi f(x) = \int \Pi(x, dx')f(x')$ for the Markov kernel $\Pi$ acting on $f$.

\subsection{Theoretical results}\label{sec:theory}

We will for brevity present this section in the Fisherian setting. By considering $\{\X, \theta\}$ as the latent variables instead of $\X$, the results are applicable also to the Bayesian setting.

Convergence of the SAEM algorithm has only been established for models in the exponential family. 
In addition to the requirements on the step size sequence in \eqref{eq:gamma}, 
the essence of the assumptions used by \cite{KuhnL:2004} are:
\begin{enumerate}[({A}1)]
	\item\label{ass:1} The parameter space $\Theta$ is an open subset of $\mathbb{R}^p$.
	The model belongs to the exponential family, and the log-likelihood function and its components $\phi, \psi$ and $S$ are sufficiently smooth, differentiable and integrable.
	\item\label{ass:2} A unique solution to the maximization problem in the (M)-step exists, and that mapping from $\mathbb{S}_k$ to $\theta_k$ is sufficiently differentiable.
	\item\label{ass:3} $\setX$ is compact and $S$ is continuous on $\setX$. 
	\item\label{ass:4} The Markov kernel $\kernel{\theta}$ for sampling $\X$ is uniformly ergodic
	uniformly in $\theta$.
	Furthermore, 
    $\kernel{\theta}$ 
	is Lipschitz continuous w.r.t.\ $\theta$ uniformly in $\X$.
\end{enumerate}

\textit{Remark:}
    For more precise statements of the actual assumptions under which we prove convergence of PSAEM, see Appendix~\ref{app:B}.

Under 
such assumptions
 \cite{KuhnL:2004} show that SAEM converges to a stationary point of the likelihood surface.
Assumption (A\ref{ass:1})-(A\ref{ass:3}) define the class of models~\eqref{eq:ssm} for which convergence is proven.
The compactness assumption on $\setX$ is strong, and 
ensures that $\mathbb{S}_k$ cannot diverge, but is
not strictly necessary. The more general case is, however, far from trivial, see \cite[Section~5]{DelyonLM:1999}, \cite{AndrieuMP:2005} and \cite{AndrieuV:2014}.
Assumption (A\ref{ass:4}) puts requirements (uniform ergodicity and Lipschitz continuity) on the MCMC kernel that is used, which is PGAS in our case. Uniform ergodicity has been shown for PGAS under a boundedness assumption on the weights of the conditional particle filter \cite[Theorem 3]{LindstenJS:2014}. 
In Appendix~\ref{app:B} we extend this result to hold uniformly in $\theta$ under assumption (A\ref{ass:6}), stated below.
What has not previously been shown, though, is Lipschitz continuity of the PGAS Markov kernel.
This property is establish below under the following additional assumption.

\begin{enumerate}[({A}1)]
	\setcounter{enumi}{4}
	\item\label{ass:6} 
	There exists constants $L_1, L_2 < \infty$, $\delta_1, \delta_2 > 0$ and $\kappa_1, \kappa_2 < \infty$, independent of $x_{t-1}, x_t, \theta$, such that, for all $x_{t-1},x_t\in \setX$ and all $t=1,\dots,T$,
	\begin{enumerate}
		\item \emph{Lipschitz continuity of transition and likelihood densities:} For all $\theta,\ttheta\in\Theta$,
		\begin{align*}
		| p_\theta(x_t \mid x_{t-1}) - p_{\ttheta}(x_t\mid x_{t-1}) | &\leq L_1\|\theta-\ttheta\|, \\
		| p_\theta(y_t \mid x_t) - p_{\ttheta}(y_t\mid x_t) | &\leq L_2\|\theta-\ttheta\|.
		\end{align*}
		\item\label{ass:6-weight-lower-bound} \emph{Strong mixing:} For all $\theta\in\Theta$,
		\(\delta_1 \leq p_\theta(x_t\mid x_{t-1}) \leq \kappa_1\) and \(\delta_2\leq p_\theta(y_t\mid x_t) \leq \kappa_2.\)
	\end{enumerate}
\end{enumerate}

\textit{Remark:} The lower bound on the state transition and likelihood functions in (A\ref{ass:6-weight-lower-bound}), commonly referred to as the \emph{strong mixing condition}, are indeed strong but have traditionally been used for establishing many theoretical results on SMC, see for instance \cite{DelMoral:2004}.
Furthermore, this assumption essentially boils down to compactness of $\setX$, which is assumed in (A\ref{ass:3}) already. The strong mixing condition has been weakened for some results \cite{Whiteley:2013,Handel:2009}, and could possibly be extended further.

\begin{theorem}[Lipschitz continuity of PGAS]\label{thm:lipschitz}
	Assume (A\ref{ass:3}) and (A\ref{ass:6}) and let $\kernel{\theta}$ denote the PGAS Markov kernel \eqref{eq:Pi-X-def}. Then there exists a constant $C<\infty$ such that for any bounded function $f : \setX^{T+1} \to \reals$, it holds that for all $\theta,\ttheta \in\Theta$,
	\begin{align*}
	\| \kernel{\theta} f - \kernel{\ttheta}f\|_\infty \leq C \|f\|_\infty \|\theta-\ttheta\|.
	\end{align*}
\end{theorem}
\begin{proof}
	See Appendix \ref{app:A}.
\end{proof}

We may now piece all results together into the main theorem of this section, which establishes the convergence of \psaem.

~

\begin{theorem}[Convergence of \psaem]\label{thm:conv}
	Assume (A\ref{ass:1})-(A\ref{ass:3}); see precise statements in Appendix~\ref{app:B}. Additionally, assume (A\ref{ass:6}) and let $\theta_k$ be computed by \cref{alg:psaemfexp}. Then, with probability 1, $\lim_{\ei\to\infty} d(\theta_{\ei},\mathcal{L}) = 0$, where $d(\theta,\mathcal{L})$ denotes the distance from $\theta$ to the set $\mathcal{L} = \{\theta\in\Theta:\frac{\partial}{\partial\theta}p_{\theta}(\Y) = 0\}$.
\end{theorem}
\begin{proof}
    The proof, together with precise statements of the assumptions, is given in Appendix~\ref{app:B}.
	The big picture is that \cref{thm:lipschitz} (together with existing ergodicity results) implies (A\ref{ass:4}), and therefore \cref{thm:conv} follows from \cite[Theorem~1]{KuhnL:2004}. 
\end{proof}

\subsection{Practical considerations}\label{sec:practice}
Even though \cref{thm:conv} gives a reassuring theoretical foundation for using \psaem, it does not give any practical advice on some of the (few) tuning parameters available: the choice of step length $\{\gamma_k\}_{k=1}^\infty$ or the number of particles $N$ in Algorithm~\ref{alg:cpfas}.

A common choice for step length is $\gamma_k = k^{-\alpha}$, and the requirements~\eqref{eq:gamma} are fulfilled for any $\alpha\in(\frac{1}{2},1]$. In our experience, it is often advisable to choose $\alpha < 1$, perhaps $\alpha=0.7$, not to constrain the steps too much. Even though not necessary, the initial convergence speed can sometimes be improved by setting some initial step lengths to constant $1$, before starting the sequence of decreasing step lengths.

For $N$, we have to make a balance between a well mixing Markov kernel (large $N$) and the computational load (small $N$). Let $K$ denote the number of iterations of \psaem, and assume that the computational budget available is such that the product $KN$ is limited. In such a situation, the general advice would be to take $N$ `small' and $K$ `large'. However, if $N$ is too small, the Markov kernel will not mix well, affecting the convergence speed. To monitor the mixing, the overlap between two consecutive state trajectories $\X[\ei-1]$ and $\X[\ei]$ could be computed, and if it exceeds a certain threshold, say $90\%$, a warning could be raised that the mixing is not sufficient and $N$ should be increased.

\section{Experiments and applications}
We will in this section first (\cref{ex:lgss}) illustrate the behavior of \psaem on a small toy example (where the maximum likelihood estimate can be found exactly), and study the advantage over a standard Monte Carlo EM implementation for the same problem. We will thereafter turn to three different applications, namely parameter estimation in a non-linear state-space model (the Fisherian setting, \cref{ex:wt}), and hyperparameter estimation (Bayesian setting) in infinite factorial dynamical models (\cref{ex:ifdm}) and Gaussian process state-space models (\cref{ex:gpssm}), respectively. Full details for all examples are found in Appendix~\ref{app:C}.

\subsection{Linear Gaussian state-space model}\label{ex:lgss}
We consider $T = 300$ data points from the model
\begin{subequations}\label{eq:ex:simple}
	\begin{align}
	x_{t+1} &= \theta x_t + w_t, \quad & w_t&\sim\mathcal{N}(0,1),\\
	y_{t} &= x_t + e_t, &e_t&\sim\mathcal{N}(0,0.3),
	\end{align}
\end{subequations}
with $\theta\in(-1,1)$. We apply \psaem and four alternative methods. A close relative to \psaem, namely PIMH-SAEM (using particle independent Metropolis--Hastings instead of PGAS; \cite{AndrieuV:2014,DonnetS:2014}), is applied. We also use two different Monte Carlo EM solutions~\eqref{eq:MCEMf}, one using the forward filter backward simulator (FFBSi) smoother\footnote{This is similar to the method proposed by \cite{SchonWN:2011}, but it uses a more efficient smoother.} \cite{OlssonDCM:2008,GodsillDW:2004} and one using the particle-based rapid incremental (PaRIS) smoother \cite{OlssonW:2017}. Compared to FFBSi, the PaRIS smoother has the computational advantage that it approximates not the entire distribution $p_{\theta_k}(\X\mid\Y)$, but only $p_{\theta_k}(x_t,x_{t+1}\mid\Y)$, which in fact is sufficient for the Fisherian problem. 
In fact, PaRIS is an online smoothing algorithm so it can also be combined with online-EM as proposed by \cite{OW:2015}. The online-EM method solves indeed also the (challenging) online problem, and is included in the comparison.
We iterate each method $1\,000$ times, and study the convergence to the true maximum likelihood estimate (which is available exactly in this toy model). For the online-EM algorithm we loop over the $T=300$ data points $1\,000$ times (note that this method makes one parameter update per single time step).
All methods are applied with different numbers of particles $N$.

\begin{figure*}
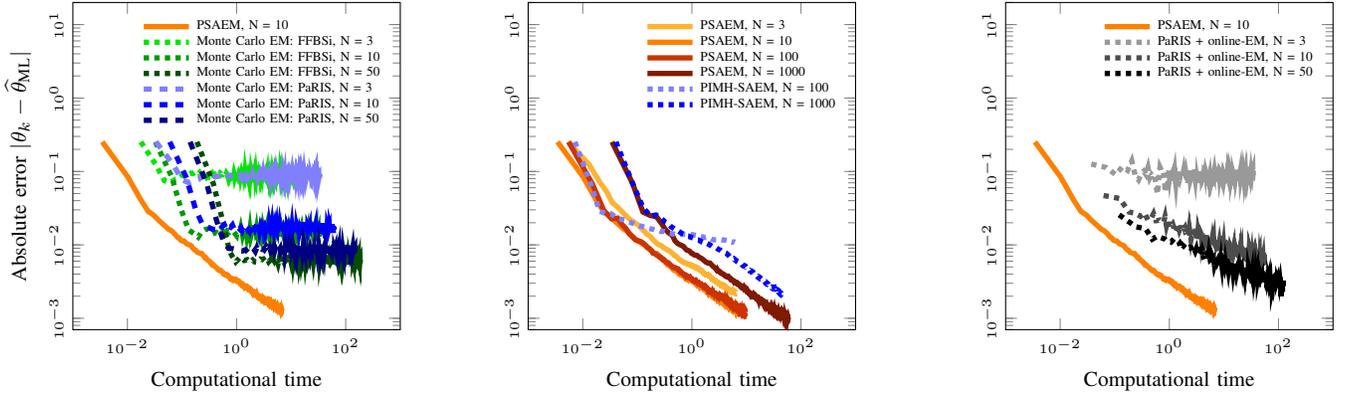

	\centering
	\begin{subfigure}[t]{.3\linewidth}
		\centering
		\input{fig/lgss1}
		\caption{Besides faster convergence (in terms of computational time), \psaem does not suffer from the bias present in Monte Carlo EM. The bias is caused by the finite number of samples in the integral \eqref{eq:MCEMf}, and the bias vanishes only as $N\to\infty$, in contrast to \psaem for which $\ei\to\infty$ is enough.}
	\end{subfigure}\hfill
	\begin{subfigure}[t]{.3\linewidth}
		\centering
		\input{fig/lgss2}
		\caption{The optimal $N$ for \psaem in this problem appears to be in the range 10--100; smaller $N$ causes poor mixing and slower convergence; larger $N$ increases computational cost without improving mixing. PIMH-SAEM with $N=100$ struggles because of poor mixing, whereas $N=1000$ mixes better at a higher computational cost.}
	\end{subfigure}\hfill
	\begin{subfigure}[t]{.3\linewidth}
		\centering
		\input{fig/lgss3}
		\caption{The convergence of online-EM with the PaRIS smoother is theoretically not fully understood, but appears for $N=3$ to suffer from a similar bias as Monte Carlo EM. For this example it appears to converge with larger $N$, however at a significantly higher computational cost than \psaem.}
	\end{subfigure}
	\caption{Estimation of $\theta$ in~\eqref{eq:ex:simple} with five different methods; the proposed \psaem,  PIMH-SAEM (similar to \psaem, but with a PIMH Markov kernel instead of PGAS), two Monte Carlo EM implementations (using the FFBSi and PaRIS smoother, respectively) and online-EM with the PaRIS smoother. The methods are run with various number of particles $N$, cf. the discussion on $N$ in Section~\ref{sec:practice}. Their average evolution of the absolute error $|\theta_\ei-\widehat{\theta}_\text{ML}|$ (over 200 runs), where $\widehat{\theta}_\text{ML}$ is the exact maximum likelihood estimate, is shown as a function of the wall clock time. All methods are run for $1\,000$ iterations, and there is a linear relationship between $k$ and computational time for all methods.}
	\label{fig:lgss}
\end{figure*}

In \cref{fig:lgss}, the evolution of the absolute error is shown as a function of computational time on the same standard desktop computer with comparable implementations, averaged over 200 realizations of each algorithm. Note that \psaem and PIMH-SAEM converge as $\ei\to\infty$ (for fixed $N$), whereas Monte Carlo EM has a non-vanishing bias which only decreases as $N\to\infty$. In other words, $\ei\to\infty$ is not sufficient for convergence in Monte Carlo EM. Comparing PSAEM and PIMH-SAEM, the latter requires a significantly larger number of particles than \psaem, and has therefore a higher computational cost. This difference is likely to be even more pronounced for larger values of $T$, due to superior scaling properties of PGAS compared to PIMH.

\subsection{Cascaded water tanks}\label{ex:wt}
We consider the benchmark problem of learning a model for a cascaded water tank system, using the data presented by\footnote{See also \url{http://www.nonlinearbenchmark.org}} \cite{SN:2017}. A training and a test data set of input-output data samples $\{u_t, y_t\}$, each with $T=1024$ data points, are provided. The data is recorded from an experimental setup where water is pumped into an upper water tank, from which it flows through a small opening into a lower water tank, and from there through another small opening into a basin. During the data collection, the tanks occasionally overflowed, and the excess water from the upper tank partially flowed into the lower tank. 
Only the pump voltage (input) and the water level in the lower tank (output) is measured each $T_s = 4$ second, and the problem is to predict the water level in the lower tank given only the pump voltage. A physically motivated discrete-time nonlinear state-space model (partly adopted from \cite{HRC+:16}) is
\vspace{-1em}

\footnotesize
	\begin{align}
	x^u_{t+1} =& 10 \wedge x_t^u + T_s(-k_1\sqrt{10 \wedge x_t^u}-k_2\{10 \wedge x_t^u\} + k_5 u_t) + w_t^u\nonumber\\
	x^l_{t+1} =& 10 \wedge x_t^l + T_s(k_1\sqrt{10 \wedge x_t^u}+k_2\{10 \wedge x_t^u\} - k_3\sqrt{10 \wedge x_t^l}\nonumber\\
	& - k_4\{10 \wedge x_t^l\} + k_6\{(x_t^u-10) \vee 0\}) + w_t^l \nonumber\\
	y_t =& 10 \wedge x^l_t + e_t,\label{eq:ex:wt}
	\end{align}%
	
\normalsize
\noindent where the states $x^u_t\in\mathbb{R}$ and $x^l_t\in\mathbb{R}$ are the water levels plus the inflow in the upper and lower tank, respectively. The parameters $k_1,k_2,k_3,k_4,k_5$ represent unknown physical quantities, such as tank and hole diameters, flow constants, pump efficiency, etc. Each tank has height $10$ (in the scale of the sensor), and $k_6$ and $10\wedge(\dots)$ is motivated by the overflow events. The initial level of the upper water tank is modeled as $x^u_0\sim\mathcal{N}(\xi_0,\sqrt{0.1})$, with $\xi_0$ unknown. Furthermore, $w_t^1$, $w_t^2$ and $e_t$ are assumed to be zero mean white Gaussian noise with unknown variances $\sigma_w^2$ and $\sigma_e^2$, respectively. All in all, the unknown parameters are $\theta = \{k_1,k_2,k_3,k_4,k_5,k_6, \sigma_e^2, \sigma_w^2, \xi_0\}$.

The model belongs to the exponential family, and we can thus apply \psaem as presented in Algorithm~\ref{alg:psaemfexp} to find a maximum likelihood estimate of $\theta$. We initialize $\theta$ randomly around physically reasonable values, and run \psaem for 50 iterations with $N=100$ (taking a few seconds on a standard desktop computer). The obtained results are reported in Table~\ref{tab:wt} together with the best performing result previously published (to the best of our knowledge). Many previously published methods use a more data-driven approach, but the relatively small amount of data makes the encoding of physical knowledge important, as done here by~\eqref{eq:ex:wt} and \psaem.

\begin{table}\footnotesize\centering
	\begin{subfigure}{.13\linewidth}\centering
		\includegraphics[width=\linewidth]{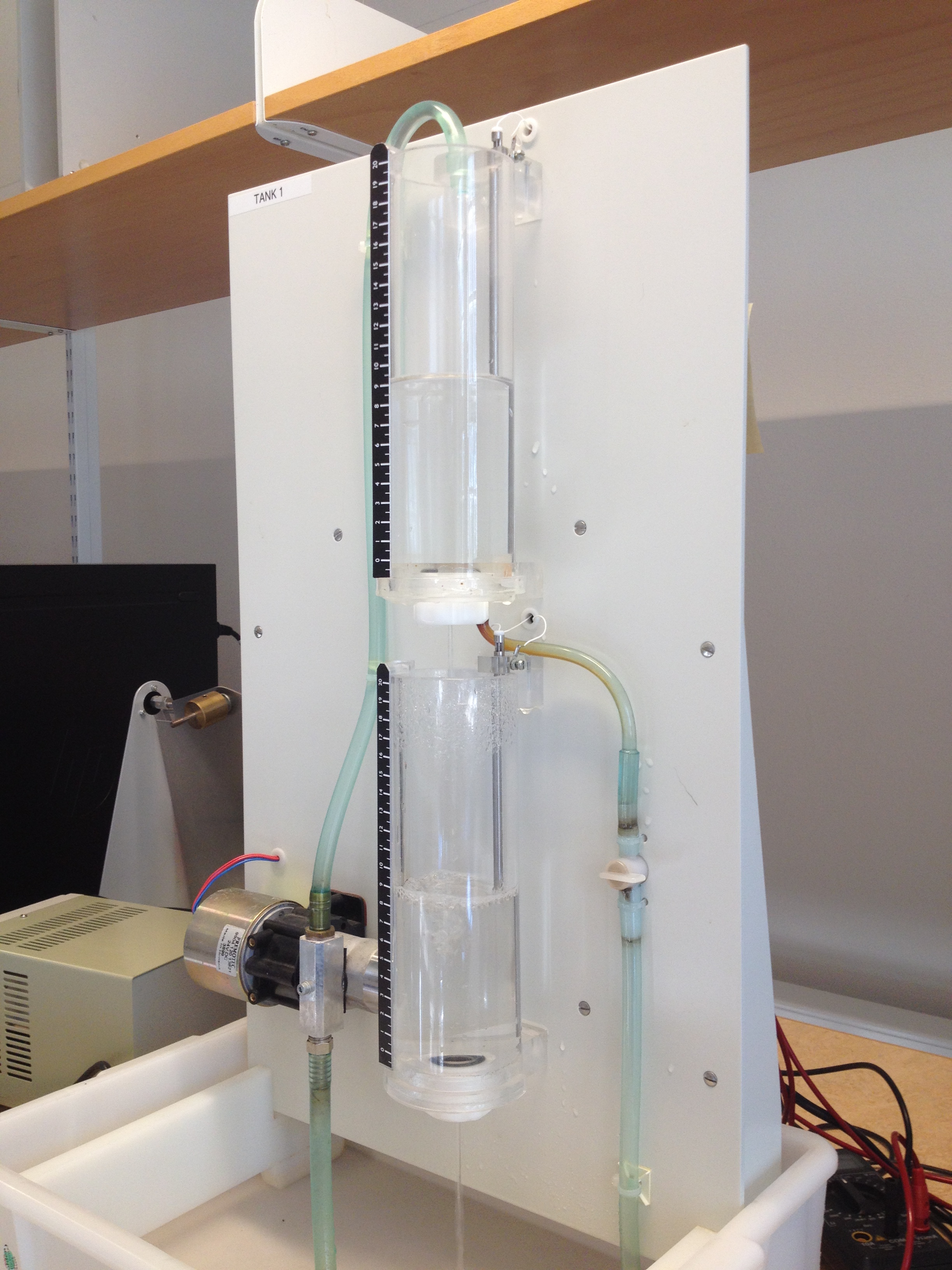}
	\end{subfigure}
	\begin{subfigure}{.8\linewidth}\centering
		\begin{tabular}{cc}
			Model & Simulation (test data) \\\midrule
			Initial model to \psaem & 2.85\\
			Estimated with \psaem & \textbf{0.29} \\
			\cite{RTM:2017} & 0.34
		\end{tabular}
	\end{subfigure}
	\caption{The cascaded water tank setup and modeling results. We initialize the 9 unknown parameters with an ad-hoc educated guess (top row), and then optimize them with \psaem (middle row). We also include the best performing result previously published (last row). The figure of merit is root-mean-squared error for simulation on the test data.}
	\label{tab:wt}
\end{table}

\subsection{Hyperparameters in infinite factorial dynamical models}\label{ex:ifdm}

The infinite factorial dynamical model (iFDM, \cite{ValeraRSP:2015}) is a Bayesian non-parametric model for separation of aggregated time-series into independent sources. By using a Markov Indian buffet process, the number of sources (dimensionality of the hidden state) does not have to be upper bounded a priori. Each source is modeled as a (discrete or continuous) Markov chain which evolves independently of the other. To solve the inference problem, i.e., performing the actual source separation, PGAS has proven useful \cite{ValeraRSP:2015}. There is, however, a multitude of hyperparameters in this Bayesian setting, and we demonstrate how the procedure by \cite{ValeraRSP:2015} easily can be extended with \psaem to automatically estimate hyperparameters on-the-fly, reducing the need for extensive manual tuning.

\begin{figure}[t]
	\centering
	\begin{tikzpicture}
	
	\begin{axis}[%
	width=.3\linewidth,
	height=.3\linewidth,
	at={(0,0)},
	scale only axis,
	axis on top,
	xmin=0.5,
	xmax=15.5,
	xtick={ 1,  2,  3,  4,  5,  6,  7,  8,  9, 10, 11, 12, 13, 14, 15},
	xlabel={Speaker index},
	y dir=reverse,
	ymin=0.5,
	ymax=1087.5,
	ytick={\empty},
	ylabel={Time},
	axis background/.style={fill=white},
	title={Ground truth},
	legend style={legend cell align=left, align=left}
	]
	\addplot [forget plot] graphics [xmin=0.5, xmax=15.5, ymin=0.5, ymax=1087.5] {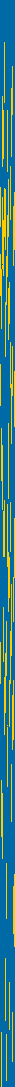};
	\end{axis}
	
	\begin{axis}[%
	width=.3\linewidth,
	height=.3\linewidth,
	at={(0.33\linewidth,0)},
	scale only axis,
	axis on top,
	xmin=0.5,
	xmax=15.5,
	xtick={ 1,  2,  3,  4,  5,  6,  7,  8,  9, 10, 11, 12, 13, 14, 15},
	xlabel={Speaker index},
	y dir=reverse,
	ymin=0.5,
	ymax=1087.5,
	ytick={\empty},
	axis background/.style={fill=white},
	title={Sample without EB},
	legend style={legend cell align=left, align=left}
	]
	\addplot [forget plot] graphics [xmin=0.5, xmax=15.5, ymin=0.5, ymax=1087.5] {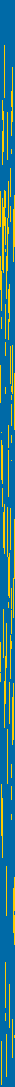};
	\end{axis}
	
	\begin{axis}[%
	width=.3\linewidth,
	height=.3\linewidth,
	at={(0.66\linewidth,0)},
	scale only axis,
	axis on top,
	xmin=0.5,
	xmax=15.5,
	xtick={ 1,  2,  3,  4,  5,  6,  7,  8,  9, 10, 11, 12, 13, 14, 15},
	xlabel={Speaker index},
	y dir=reverse,
	ymin=0.5,
	ymax=1087.5,
	ytick={\empty},
	axis background/.style={fill=white},
	title={Sample with EB},
	legend style={legend cell align=left, align=left}
	]
	\addplot [forget plot] graphics [xmin=0.5, xmax=15.5, ymin=0.5, ymax=1087.5] {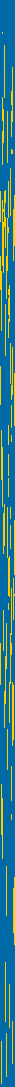};
	\end{axis}
	\end{tikzpicture}%
	\caption{The cocktail problem, introduced by \cite{GaelTG:2009,ValeraRSP:2015}, amounts to inferring the number of speakers (columns) and their periods of talking and being quiet (yellow and blue, respectively) from an aggregated observation (not shown) during a time sequence (y-axis). The middle panel is a sample from the solution by \cite{ValeraRSP:2015}, and the right panel a sample after we have extended that solution with \psaem to automatically estimate some hyperparameters. Both solutions infer the correct number of speakers (15), but the solution with \psaem is slightly less prone to switch between quiet and talking (closer to the ground truth, left). However, the main advantage of \psaem, not present in the plot itself, is the lessened need for manual tuning of hyperparameters at almost no extra computational cost.}
	\label{fig:ifdm}
\end{figure}

We will consider the cocktail party problem originating from \cite{GaelTG:2009}, to which iFDM has been applied \cite[Section 4]{ValeraRSP:2015}. The voices from 15 different speakers is aggregated into a $T=1085$ long sequence, together with some noise, and the problem is to jointly infer (i) the number of speakers (dimension of $x_t$), (ii) when each speaker is talking (the trajectory $x_t$) and (iii) the dynamics of each speaker (how prone s/he is to talk). 
Each speaker is modeled as a Markov chain with two states, `talking' or `quiet', and the posterior distribution over its transition probabilities is inferred individually for each speaker. The Beta distribution is used as prior for these probabilities, and the hyperparameters for the Beta distribution are manually chosen by \cite{ValeraRSP:2015}. We outline in \cref{alg:ifdm} how the inference procedure can be extended with \psaem (new lines are marked with blue). In addition to the lessened burden of manual hyperparameter tuning, we can also report slightly improved results: With the hyperparameters automatically found by \psaem, the average number of switches between `quiet' and `talking' in posterior samples are closer to ground truth (84 instead of 86, ground truth: 62) and the average value of the complete data likelihood of the posterior samples increases. Of course, \psaem could be applied also to other hyperparameters in the problem, following the very same pattern. Posterior samples of $x_t$ are shown in \cref{fig:ifdm}.

\begin{algorithm}[t]\algfontsize
	\caption{\psaem for infinite factorial dynamical models: the cocktail party example}
	\begin{algorithmic}[1]
		\For{$\ei = 0$ to $K$}
		\State Update the state dimensionality (number of speakers) $M_+$ using slice sampling.
		\State Sample a state trajectory using PGAS.
		\State Gibbs update of the transition probabilities $\{b^m\}_{m=1}^{M_+}$ for each speaker.
		\State \textcolor{blue}{$\saS_{\ei} \leftarrow (1-\gamma_\ei)\saS_{\ei-1} + \gamma_\ei S(\{b^m\}_{m=1}^{M_+})$, with $S$ being sufficient statistics for the Beta distribution.}
		\State \textcolor{blue}{$\eta_{\ei}\leftarrow \argmax_{\eta} \saemQb_{\ei}(\eta) = \argmax_{\eta} \langle \saS_\ei,\phi(\eta)\rangle-\psi(\eta)$.}
		\State Gibbs update of noise variance parameters.
		\EndFor
	\end{algorithmic}
	\label{alg:ifdm}
\end{algorithm}

\subsection{Hyperparameter estimation in Gaussian process state-space models}\label{ex:gpssm}

Gaussian process state-space models are a combination of the state-space model and the Gaussian process (GP) model as

\vspace{-1em}

\footnotesize
\begin{subequations}
	\begin{align}
	x_{t+1} &= f(x_t) + w_t, &&f\sim\GP{m^f_\eta,K^f_\eta}, &&w_t \sim \N{0,\Sigma_n^f},\\
	y_{t} &= g(x_t) + e_t, &&g\sim\GP{m^g_\eta,K^g_\eta}, &&e_t \sim \N{0,\Sigma_n^{g}},
	\end{align}
\end{subequations}

\normalsize \noindent
or variations thereof. As in any state-space model, only $\Y$ is observed and not $\X$, and standard GP regression methods \cite{RasmussenW:2006} can therefore not be used to learn the posterior $p_\eta(f,g\mid y_{1:T})$. Consequently, learning of the GP hyperparameters $\eta$---usually done via empirical Bayes $\etaml = \argmax_{\eta}p_\eta(f,g\mid y_{1:T})$---is not straightforward either.

Despite the computational challenges, it has been argued that the model is versatile and powerful by its combination of the dynamic state-space model and the nonparametric and probabilistic GP, and has for this reason achieved attention in the machine learning literature. One proposed solution is to use PGAS for learning the model \cite{SSS+:2016,FrigolaLSR:2013,FrigolaLSR:2014}, and we extended that solution with \psaem to also include estimation of the hyperparameters at almost no extra computational cost.

\begin{algorithm}[t]\algfontsize
	\caption{PSAEM for hyperparameter estimation in Gaussian process state-space models}
	\begin{algorithmic}[1]
		\State Initialize ${x}_{1:T}[0],\theta[0],\eta_0$.
		\For{$\ei = 0$ to $K$}
		\State\label{alg:gp-ssm:cpfas} Sample ${\X[\ei]}{\:\big\vert\:} {\theta}_{\ei-1}, \eta_{\ei-1}$ using Algorithm~\ref{alg:cpfas}.
		\State\label{eq:dummy} Sample ${\theta}_{\ei}{\:\big\vert\:} \eta_{\ei-1}, \X[\ei]$ with a closed-form expression.
		\State\label{alg:gp-ssm:psaem1} \textcolor{blue}{Update ${\saS_{\ei}}\leftarrow  (1-\gamma_k)\saS_{\ei-1} + \gamma_k S(\theta[\ei])$.}
		\State\label{alg:gp-ssm:psaem2} \textcolor{blue}{Solve ${\eta_{\ei}}\leftarrow \argmax_{\eta} \saemQb_{\ei}(\eta)$,\newline $\saemQb_{\ei}(\eta) = \argmax_{\eta} \langle \saS_\ei,\phi(\eta)\rangle-\psi(\eta)$.}
		\EndFor
	\end{algorithmic}
	\label{alg:gp-ssm}
\end{algorithm}

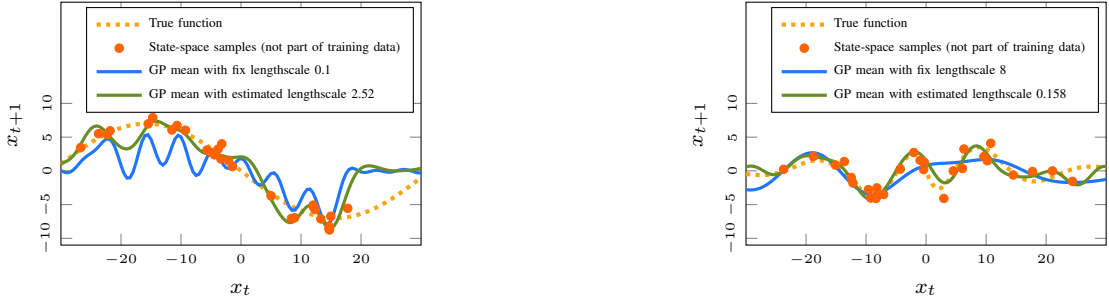
\begin{figure*}
	\centering
	\begin{subfigure}{.33\linewidth}
		\input{fig/GPi}
	\end{subfigure}\hspace{3cm}
	\begin{subfigure}{.33\linewidth}
		\input{fig/GPii}
	\end{subfigure}	
	\caption{Estimation of hyperparameters in Gaussian process state-space models. Two different examples, with different data, are shown. Since the states $x_t$ are unobserved in the GP-SSM, the learning is more challenging than standard GP regression. We have extended an PGAS procedure \cite{SSS+:2016} to also include estimation of the hyperparameters with \psaem. The blue line is the posterior mean of $p(f\mid y_{1:T},\eta)$ with $\eta$ being a fixed lengthscale, whereas the length scale $\eta$ is estimated with \psaem for the green line. The true function is dashed yellow, alongside with the $T = 40$ samples $\X$ (orange dots) underlying the training data (only $\Y$, not shown, is available when learning the model). The main point of this example is a proof of concept for hyperparameter estimation using \psaem in the challenging GP-SSM model.}
	\label{fig:gp-ssm}
\end{figure*}

We consider the solution proposed by \cite{SSS+:2016}, in which the nonparametric GP is approximated with a reduced-rank representation with a finite parameter set $\theta$. We introduce \psaem for this solution in Algorithm~\ref{alg:gp-ssm} (new lines in blue). Since the computational burden in practice is dominated by running the conditional particle filter, the inclusion of \psaem adds very little extra computational cost. An example of estimation of the length scale in a Gaussian process state-space model is shown in Figure~\ref{fig:gp-ssm}, where the space of $x_t$ is one-dimensional and $g(x) = x$ is considered known, but the noise level is significant with $\sigma_n^f = \sigma_n^{g} = 1$.

\section{Conclusions}

We have presented \psaem for learning nonlinear state-space models, both in a maximum likelihood and an empirical Bayes setting. We have also summarized the available theoretical results, and added a missing piece about continuity of the PGAS Markov kernel in order to show convergence. Besides maximum likelihood parameter estimation, we believe \psaem has a great potential also for models where PGAS is currently used and automatic estimation of hyperparameters via \psaem can be achieved with only a small modification to existing implementations.


%

\appendices

\section{Proof of \cref{thm:conv}, convergence of PSAEM}\label{app:B}

We first list the assumptions behind \cref{thm:conv} in more detail.
First, assumption (A\ref{ass:1}) is explicitly:
\begin{itemize}
    \item $\Theta$ is an open set. The model belongs to the exponential family \eqref{eq:expf2}, where $\phi$ and $\psi$ are twice differentiable on $\Theta$ and $S$ is a Borel function in $\X$ taking its values in an open subset $\mathcal{S}$ of $\mathbb{R}^\ell$. The convex hull of $S(\mathbb{R}^{m\times(T+1)})$ is included in $\mathcal{S}$. Furthermore, for all $\theta\in\Theta$, $\int |S(\X,\Y)|p_\theta(\X\mid \Y)d\X<\infty$ and $\int S(\X,\Y)p_\theta(\X\mid \Y)d\X$ is continuously differentiable \wrt $\theta$. 
    \item The log-likelihood function $\log p_\theta(\Y)$ is continuously differentiable once and $\ell$ times differentiable on $\Theta$. Furthermore, $\partial_\theta \int p_\theta(\X,\Y) d\X = \int \partial_\theta p_\theta(\X,\Y) d\X$.
\end{itemize}
This assumption corresponds to assumptions (M1), (M2), (M3) and  
(M4) of \cite{KuhnL:2004}. Furthermore, $\ell$ times differentiability of the log-likelihood function corresponds to part of their assumption (SAEM2). 

Our assumption (A\ref{ass:2}), which corresponds to (M5) and the remaining part of (SAEM2) of \cite{KuhnL:2004}, is more explicitly:
\begin{itemize}
    \item A unique solution to the maximization problem in the (M)-step exists, and that mapping from $\mathbb{S}_k$ to $\theta_k$ is continuously differentiable once and $\ell$ times differentiable. 
\end{itemize}
Our assumption (A\ref{ass:3}) is:
\begin{itemize}
    \item $\setX$ is compact and $S$ is continuous on $\setX$.
\end{itemize}
Since $S$ is a continuous function on a compact subset of $\reals^m$, it is also bounded.
This corresponds to \cite[(SAEM3')1, (SAEM3')4]{KuhnL:2004}.
Furthermore, since $S$ is continuous and $\setX$ is compact, the image $S(\setX)$ is also compact, and so is its convex hull. It follows that the sequence $\{\mathbb{S}_k\}_{k\geq 1}$ takes its values in a compact subset of $\mathcal{S}$, which is \cite[(C)]{KuhnL:2004}.

Finally, since our design choice \eqref{eq:gamma} corresponds to \cite[(SAEM1)]{KuhnL:2004}, the only missing pieces in order to apply \cite[Theorem 1]{KuhnL:2004} are \cite[(SAEM3')2-3]{KuhnL:2004}, which have to do with Lipschitz continuity and uniform ergodicity of the PGAS kernel; see our (A\ref{ass:4}). 

Consider first ergodicity. We require uniform ergodicity, uniformly in $\theta$ \cite[(SAEM3')3]{KuhnL:2004}.
Specifically,
for any bounded function $f : \setX^{T+1} \to \reals$, let
$\bar f_{\theta}(x_{0:T}) = f(x_{0:T}) - \int f(x_{0:T}) p_{\theta}(x_{0:T}\mid y_{0:T}) \myd x_{0:T}$. Then, we require,
\begin{align*}
    \sup_{\theta\in\Theta}\| \kernel{\theta}^{j} \bar f_\theta  \|_\infty \leq M\rho^j \|f\|_\infty
\end{align*}
for all $j\geq 0$ and for constants $M < \infty$ and $\rho \in [0,1)$ independent of $\theta$.

It follows from \cite[Theorem~1]{LindstenDM:2015} that the PGAS Markov kernel $\kernel{\theta}$ satisfies a global Doeblin condition and that it is uniformly ergodic, such that,
\begin{align*}
	\| \kernel{\theta}^{j} \bar f_\theta  \|_\infty \leq \rho_\theta^j \|f\|_\infty
\end{align*}
for $j\geq 0$ and where
\begin{align*}
    \rho_\theta = 1 - \prod_{t=0}^T \frac{N-1}{2B_{t,T}^\theta + N - 2}
\end{align*}
and where the terms $B_{t,T}^\theta$ are defined in \cite[Eq.~10]{LindstenDM:2015}.
It remains to prove that these terms are bounded uniformly in $\theta$. However, under the strong mixing assumption (A\ref{ass:6}), we have from the proof of \cite[Proposition~5]{LindstenDM:2015} that $B_{t,T}^\theta \leq (\kappa_1\kappa_2)/(\delta_1\delta_2)$. This proves the first part of our (A4), which is the same as \cite[(SAEM3')3]{KuhnL:2004}.


The final ingredient is the Lipschitz continuity of the Markov kernel, corresponding to \cite[(SAEM3')2]{KuhnL:2004}. This corresponds to our \cref{thm:lipschitz} (proven in Appendix~\ref{app:A}).
A slight difference between our \cref{thm:lipschitz} and \cite[(SAEM3')2]{KuhnL:2004}, however, is that 
the latter
assumes that the Markov transition kernel admits a density with respect to Lebesgue measure and that this density function is Lipschitz continuous. 
Our continuity result is instead expressed in terms of total variation distance.
The condition (SAEM3')2 is used by \cite{KuhnL:2004} to prove their Lemma~2, see \cite[p.~129]{KuhnL:2004}. 
Thus, to complete the picture we provide a lemma which replaces \cite[Lemma~2]{KuhnL:2004}. The result---which extends the continuity of the PGAS Markov kernel to the $k$-fold kernel---is a special case of \cite[Proposition~B.2]{AndrieuMP:2005}, but for completeness we repeat the proof here.
\begin{lemma}\label{lem:1}
	Assume that the conditions of \cref{thm:lipschitz} hold. Then, there exists a constant $D\leq\infty$ such that for any $k\geq 0$ and any bounded function $f$,
	\begin{align*}
	\| \kernel\theta^k f - \kernel{\ttheta}^k f\|_\infty \leq D \|f\|_\infty \|\theta-\ttheta\|.
	\end{align*}
\end{lemma}
\begin{proof}
	Define $\bar f_{\ttheta}(x_{0:T}) = $\\$f(x_{0:T}) - \int f(x_{0:T}) p_{\ttheta}(x_{0:T}\mid y_{0:T}) \myd x_{0:T}$. Since $\bar f_\ttheta$ differs from $f$ by a constant (depending on $\ttheta$) we can write,
	\begin{multline*}
	\| \kernel\theta^k f - \kernel{\ttheta}^kf\|_\infty
	= \| \kernel\theta^k \bar f_{\ttheta} - \kernel{\ttheta}^k \bar f_{\ttheta}\|_\infty
	\leq \\\sum_{j=1}^{k} \| \kernel\theta^{k-j}(\kernel\theta-\kernel\ttheta)\kernel{\ttheta}^{j-1} \bar f_{\ttheta}\|_\infty.
	\end{multline*}
	We have,
	\begin{align*}
	\|& \kernel\theta^{k-j}(\kernel\theta-\kernel\ttheta)\kernel{\ttheta}^{j-1} \bar f_{\ttheta}\|_\infty \\ &= \sup_{x_{0:T}} \left|\int \kernel\theta^{k-j}(x_{0:T}, dx_{0:T}^\star) (\kernel\theta-\kernel\ttheta)\kernel{\ttheta}^{j-1} \bar f_{\ttheta}(x_{0:T}^\star)\right| \\
	&\leq \sup_{x_{0:T}} \int \kernel\theta^{k-j}(x_{0:T}, dx_{0:T}^\star) \left|(\kernel\theta-\kernel\ttheta)\kernel{\ttheta}^{j-1} \bar f_{\ttheta}(x_{0:T}^\star)\right| \\
	&\leq \sup_{x_{0:T}} \int \kernel\theta^{k-j}(x_{0:T}, dx_{0:T}^\star) \|(\kernel\theta-\kernel\ttheta)\kernel{\ttheta}^{j-1} \bar f_{\ttheta}\|_\infty \\
	&=\|(\kernel\theta-\kernel\ttheta)\kernel{\ttheta}^{j-1} \bar f_{\ttheta}\|_\infty.
	\end{align*}
	Now, consider the function $\kernel{\ttheta}^{\ell} \bar f_{\ttheta}(x_{0:T})$ for some $\ell \geq 0$. Recall that $\bar f_{\ttheta}$ is centered around the posterior expectation of $f$ with respect to $p_{\ttheta}(x_{0:T}\mid y_{1:T})$, which is the limiting distribution of $\kernel{\ttheta}$. Thus, by uniform ergodicity of $\kernel{\ttheta}$
	uniformly in $\ttheta$,
	\begin{align*}
	\sup_{\ttheta\in\Theta}\| \kernel{\ttheta}^{\ell} \bar f_{\ttheta} \|_\infty \leq M\rho^\ell \|f\|_\infty
	\end{align*}
	for some constants $M<\infty$ and $\rho < 1$. Consequently, the function $\kernel{\ttheta}^{\ell} \bar f_{\ttheta}(x_{0:T})$ satisfies the conditions of \cref{thm:lipschitz} and thus
	\begin{align*}
	\|(\kernel\theta-\kernel\ttheta)\kernel{\ttheta}^{j-1} \bar f_{\ttheta}\|_\infty
	\leq CM\rho^{j-1}\|\theta-\ttheta\|.
	\end{align*}
	Plugging this into the expressions above completes the proof.
\end{proof}

From this, the results of Lemma 2 in \cite{KuhnL:2004} follows for our assumptions, and hence also Theorem 1 of \cite{KuhnL:2004} and, ultimately, \cref{thm:conv} of this article.

\section{Proof of \cref{thm:lipschitz}, Lipschitz continuity of PGAS}\label{app:A}
This appendix contains a proof of \cref{thm:lipschitz}. It is based on the construction of a coupling between the Markov kernels $\kernel\theta$ and $\kernel{\ttheta}$. A similar technique has previously been used by \cite{ChopinS:2015} to prove uniform ergodicity of the Particle Gibbs kernel. An explicit coupling of conditional particle filters is used by \cite{JacobLS:2017} to construct (practical) algorithms for, among other things, likelihood estimation and unbiased estimates of smoothing functionals. 

We first review some basic properties of couplings and total variation. 
Let $P$ and $Q$ be two probability measures with densities $p$ and $q$, respectively, with respect to some reference measure $\lambda$.
Let $\mathcal{C}$ be the set of couplings of $P$ and $Q$, that is, joint probability measures with marginals $P$ and $Q$. We can then write the total variation distance between $P$ and $Q$ in the following equivalent ways:
\begin{subequations}
	\begin{align}
	\| P- Q \|_{\text{TV}} 
	&= \frac{1}{2}\sup_{|f|\leq 1}|Pf-Qf| \label{eq:TV:f} \\
	&= \lambda(\max\{p-q, 0\}) \label{eq:TV:max} \\
	&= 1-\lambda(\min\{p,q\}) \label{eq:TV:min} \\
	&= \inf_{\xi \in \mathcal{C}} \iint \I(x\neq y)\xi(dx,dy). \label{eq:TV:coupling}
	\end{align}
\end{subequations}
Note also that it is possible to explicitly construct a coupling attaining the infimum in \eqref{eq:TV:coupling}: let $\alpha = \lambda(\min\{p,q\})$, $\nu(dx) = \alpha^{-1}\min\{p(x),q(x)\}\lambda(dx)$, and
\begin{multline}
\label{eq:maximal-coupling}
\xi(dx,dy) = \alpha\nu(dx)\delta_x(dy) + \\(1-\alpha)^{-1}(P(dx)-\alpha\nu(dx))(Q(dy)-\alpha\nu(dy)).
\end{multline}
A coupling $\xi$ which attains the infimum, or equivalently which maximizes the probability of $X$ and $Y$ being identical when $(X,Y)\sim \xi$, is referred to as a \emph{maximal coupling}. 
Finally, for a coupling $\xi$, the quantity $\iint \I(x = y)\xi(dx,dy)$---that is, the probability that $X$ and $Y$ are identical under $\xi$---is referred to as the \emph{coupling probability} under $\xi$.

Now, to prove the Lipschitz continuity of the PGAS Markov kernel as stated in \cref{thm:lipschitz} we will construct a coupling $\xi_{\theta,\ttheta}(x'_{0:T}, dx^\star_{0:T}, d\widetilde x^\star_{0:T})$ of the Markov kernels $\kernel\theta(x'_{0:T}, dx^\star_{0:T})$ and $\kernel{\ttheta}(x'_{0:T}, d\widetilde x^\star_{0:T})$. 
This coupling is defined via Algorithm~\ref{alg:ccpf}, which takes $x'_{0:T}$ as input and produces $x^\star_{0:T}$ and $\widetilde x^\star_{0:T}$ as outputs, such that the marginal distributions of the output trajectories are  $\kernel\theta(x'_{0:T}, dx^\star_{0:T})$ and $\kernel{\ttheta}(x'_{0:T}, d\widetilde x^\star_{0:T})$, respectively.
For ease of notation in Algorithm~\ref{alg:ccpf}, we write $\M{P}{Q}$ for any maximal coupling (for instance the one given by \eqref{eq:maximal-coupling}) of some distributions $P$ and $Q$. 
For brevity, we also write $\M{\{p_i\}_{i=1}^N}{\{q_i\}_{i=1}^N}$ for a maximal coupling between the two discrete distributions on $\{1,\dots, N\}$ with probabilities $P(\{j\}) = p_j/\sum_{i=1}^N p_i$ and $Q(\{j\}) = q_j/\sum_{i=1}^N q_i$.

\begin{algorithm}\algfontsize
	\begin{algorithmic}[1]
		\Statex \textbf{Input:} Conditional trajectory $\X'$, parameters $\theta$ and $\ttheta$.
		\Statex \textbf{Output:} Trajectories $\X^\star$ and $\tX^\star$.
		\State Draw $x_0^i \sim p(x_0)$ and set $\tx_0^i \leftarrow x_0^i$, $i = 1, \dots, N-1$.
		\State Set $x_0^N \leftarrow x'_0$ and $\tx_0^N \leftarrow x'_0$.
		\State Set $w_0^i \leftarrow 1$ and $\tw_0^i \leftarrow 1$, $i = 1, \dots, N$.
		\For{$t = 1, 2, \dots, T$}
		\State \label{line:M-res}
		For $i = 1, \dots, N-1$, draw \[(a_t^i, \ta_t^i) \sim \M{\{w_{t-1}^j\}_{j=1}^N}{\,\{\tw_{t-1}^j\}_{j=1}^N}.\]
		\State \label{line:M-prop} For $i = 1, \dots, N-1$, draw \[(x_t^i, \tx_t^i) \sim \M{p_\theta(\,\cdot\mid x_{t-1}^{a_t^i})}{ \, p_{\ttheta} (\,\cdot \mid \tx_{t-1}^{\ta_t^i})}.\] 
		\State 
		Draw \label{line:M-as} \[\hspace{-1em}(a_t^N,\ta_t^N)\sim \M{ \{ w_{t-1}^j p_\theta(x'_t\mid x_{t-1}^j) \}_{j=1}^N}{\, \{ \tw_{t-1}^j p_\ttheta(x'_t\mid \tx_{t-1}^j) \}_{j=1}^N }.\]
		\State Set $x_t^N \leftarrow x_t'$ and $\tx_t^N \leftarrow x_t'$
		\State Set $w_t^i \leftarrow p_\theta(y_t\mid x_t^i)$ and $\tw_t^i \leftarrow p_\ttheta(y_t\mid \tx_t^i)$ for $i = 1, \dots, N$.
		\EndFor
		\State\label{line:M-J}
		Draw $(J,\tJ)\sim \M{\{ w_{T}^i\}_{i=1}^N}{\, \{ \tw_{T}^i\}_{i=1}^N }$.
		\State Set $x^\star_T = x^J_T$ and $\tx^\star_T = \tx^\tJ_T$.
		\For{$t=T-1, T-2, \dots, 0$}
		\State Set $J \leftarrow a^J_{t+1}$ and $\tJ \leftarrow \ta^\tJ_{t+1}$.
		\State Set $x^\star_t \leftarrow x^{J}_t$ and $\tx^\star_t \leftarrow \tx^{\tJ}_t$.
		\EndFor
	\end{algorithmic}
	\caption{Coupled conditional particle filters with ancestor sampling defining $\xi_{\theta,\ttheta}$.}
	\label{alg:ccpf}
\end{algorithm}

Note that for any bounded function $f$,

\vspace{-1em}

\small
\begin{multline*}
\| \kernel{\theta} f - \kernel{\ttheta}f\|_\infty 
\leq \|f\|_\infty \sup_{x_{0:T}'} \sup_{|g|\leq 1} 
| \kernel{\theta} g(x_{0:T}') - \kernel{\ttheta}g(x_{0:T}') | \\
\leq 2\|f\|_\infty \sup_{x_{0:T}'} \iint \I(x^\star_{0:T} \neq \widetilde x^\star_{0:T}) \xi_{\theta,\ttheta}(x'_{0:T}, dx^\star_{0:T}, d\widetilde x^\star_{0:T})  = 
\end{multline*}
\begin{equation*}
=  2\|f\|_\infty \sup_{x_{0:T}'} \left(1- \iint \I(x^\star_{0:T} = \widetilde x^\star_{0:T}) \xi_{\theta,\ttheta}(x'_{0:T}, dx^\star_{0:T}, d\widetilde x^\star_{0:T}) \right)
\end{equation*}

\normalsize\noindent
where we have used \eqref{eq:TV:f} and \eqref{eq:TV:coupling} for the first and second lines, respectively.
Hence, it  is sufficient to show that
\begin{align}
\label{eq:xi-coupling-proba}
\iint \I(x^\star_{0:T}= \widetilde x^\star_{0:T}) \xi_{\theta,\ttheta}(x'_{0:T}, dx^\star_{0:T}, d\widetilde x^\star_{0:T}) \geq 1-\frac{C}{2}\|\theta-\ttheta\|,
\end{align}
where $C$ is the same constant as in the statement of the theorem.

Let $\alpha_{t-1}$ denote the coupling probability for the coupling at line~\ref{line:M-res} of \cref{alg:ccpf} (and thus $\alpha_T$ is the coupling probability on line~\ref{line:M-J}). On the set $\{ x_t^{1:N} = \tx_t^{1:N}\}$ we have by \eqref{eq:TV:min}

\vspace{-1em}

\small
\begin{align}
\nonumber
\alpha_t = \sum_{i=1}^N \min\left\{ \frac{w_t^i}{\sum_k w_t^k}, \frac{\tw_t^i}{\sum_k \tw_t^k} \right\}
&\geq \frac{\sum_{i=1}^N\min\left\{ w_t^i, \tw_t^i \right\}}{\sum_{i=1}^N \max\left\{ w_t^i, \tw_t^i \right\}}  \\ 	
\geq \frac{\sum_{i=1}^N \left( \max\left\{ w_t^i, \tw_t^i \right\} - L_2\|\theta-\ttheta\| \right)}{\sum_{i=1}^N \max\left\{ w_t^i, \tw_t^i \right\}}
&\geq 1-\frac{L_2}{\delta_2}\|\theta-\ttheta\|,\label{eq:alpha-bound}
\end{align}

\normalsize\noindent
where we have used the Lipschitz continuity of the likelihood (A\ref{ass:6}a) for the penultimate inequality, and the lower bound on the likelihood (A\ref{ass:6}b) for the last inequality.

Similarly, let $\beta_t$ denote the coupling probability for the coupling on line~\ref{line:M-as}. Under assumption (A\ref{ass:6}), the product $p_\theta(y_{t-1}\mid x_{t-1}) p_\theta(x_{t} \mid x_{t-1})$ (which constitutes the unnormalized ancestor sampling weights) is bounded from below by $\delta_1\delta_2$. The product is also Lipschitz continuous in $\theta$: since
\(
|ab-cd| = |ab - ad + ad - cd| \leq |a||b-d| + |d||a-c|
\)
we have
\begin{multline*}
| p_\theta(y_{t-1}\mid x_{t-1}) p_\theta(x_{t} \mid x_{t-1}) - p_\ttheta(y_{t-1}\mid x_{t-1}) p_\ttheta(x_{t} \mid x_{t-1}) | \\ \leq (\kappa_1 L_2 + \kappa_2 L_1)\| \theta-\ttheta\|.
\end{multline*}
Therefore, on the set $\{x_{t-1}^{1:N} = \tx_{t-1}^{1:N}\}$, we have by a computation analogous to above,
\begin{align}
\label{eq:beta-bound}
\beta_t &\geq 1-\frac{\kappa_1 L_2 + \kappa_2 L_1}{\delta_1\delta_2}\|\theta-\ttheta\|.
\end{align}

Finally, let $\gamma_t^i$ denote the coupling probability for the coupling at line~\ref{line:M-prop}, for the $i$th particle. By \eqref{eq:TV:max} and \eqref{eq:TV:coupling} we have, on the set $\{ x_{t-1}^{1:N} = \tx_{t-1}^{1:N}, a_t^{1:N} = \ta_{t}^{1:N} \}$,
\begin{align}
\nonumber
\gamma_t^i &=  1 - \lambda(\max\{ p_\theta(\cdot \mid x_{t-1}^{a_t^i}) - p_\ttheta(\cdot \mid x_{t-1}^{a_t^i}), 0\}) \\
&\geq 1- L_1 \lambda(\setX) \| \theta-\ttheta\|, \label{eq:gamma-bound}
\end{align}
where $\lambda$ denotes Lebesgue measure and where the inequality follows by (A\ref{ass:6}a). By (A\ref{ass:3}), $\lambda(\setX) < \infty$. Note that the bound on $\gamma_t^i$ is independent of $i$.

Let $D = \max\{\frac{L_2}{\delta_2}, \frac{\kappa_1 L_2 + \kappa_2 L_1}{\delta_1\delta_2}, L_1 \lambda(\setX)\}$. Consider first the case
$\|\theta-\ttheta\|  \leq D^{-1}$, by which all the bounds in \eqref{eq:alpha-bound}, \eqref{eq:beta-bound}, \eqref{eq:gamma-bound} are nonnegative.
Thus, if we write $\Prb$ for probability with respect to the random variables generated by Algorithm~\ref{alg:ccpf}, we can crudely bound \eqref{eq:xi-coupling-proba} by 
\vspace{-1em}

\small
\begin{multline*}
\Prb( \{ x_t^{1:N} = \tx_{t}^{1:N}, a_t^{1:N} = \ta_t^{1:N} : t=1,\dots,T\}, J = \tJ) \\ \geq \E\left[\alpha_T\prod_{t=1}^T \left( \beta_t \prod_{i=1}^{N-1} \alpha_t \gamma_t^i \right) \right] \\
\geq \left( 1-\frac{L_2}{\delta_2}\|\theta-\ttheta\| \right)^{T(N-1)+1}
\times \left( 1-\frac{\kappa_1 L_2 + \kappa_2 L_1}{\delta_1\delta_2}\|\theta-\ttheta\| \right)^T\\
\times \left( 1- L_1 \lambda(\setX) \| \theta-\ttheta\| \right)^{T(N-1)} 
\geq (1-D\|\theta-\ttheta\|)^{2T(N-1)+T+1} \\
\geq 1-D(2T(N-1)+T+1)\|\theta-\ttheta\|,
\end{multline*}
%
\normalsize\noindent
where the last line follows from Bernoulli's inequality. 
However, since the probability is trivially bounded from below by 0, the bound above holds also for the case $\|\theta-\ttheta\| > D^{-1}$.
%
Hence, \eqref{eq:xi-coupling-proba} holds with $C=2D(2T(N-1)+T+1)$,
which proves~\cref{thm:lipschitz}.

It is worth commenting on the fact that the Lipschitz constant $C$ identified above increases with $N$, which might seem counterintuitive. However, this is an artefact of the proof technique, which is based on bounding the probability of a complete coupling of \emph{all} particles and ancestor weights generated by \cref{alg:ccpf}, which is a much stronger requirement than coupling the output trajectories only. Indeed, we expect that the Lipschitz constant stabilizes as $N\rightarrow\infty$ as the distribution of the output trajectories then converges to the joint smoothing distribution.

\section{Details about experiments}\label{app:C}
This section contains additional details regarding the experiments in Section 6.

\subsection*{Experiment 6.1--Linear Gaussian state-space model}
The step length in PSAEM, as well as PIMH-SAEM, is choosen as $\gamma_k = {k^{-0.99}}$. PSEM is implemented as a particle filter with $N$ particles and a backward simulator \cite{GodsillDW:2004} with $N$ backward trajectories. The sufficient statistics, as derived by for instance, \cite{GhahramaniH:1996}, are $\frac{1}{T}\sum_t x_tx_t^T$ and $\frac{1}{T}\sum_t x_{t-1}x_t^T$, and the maximization problem can be solved analytically.

\subsection*{Experiment 6.2--Cascaded water tanks}
The step length in PSAEM is choosen as $\gamma_{k} = 1$ for $k = 1, \dots, 30$, and $\gamma_k = (k-30)^{-0.7}$ for $k = 31, \dots$. The initial parameter values are initialized randomly around $k_1 = k_2 = k_3 = k_4 = 0.05$, $k_5 = k_6 = 0$, $\sigma^2_e = \sigma^2_w = 0.1$, $\xi_0 = 6$, and a slight $L_2$-regularization (corresponding to a $\mathcal{N}(0,10^3)$ prior) is used for $k_4$ to avoid problems if the state trajectory contains no overflow events in the lower tank. The sufficient statistics for a model on the form
\begin{align}
x_{t+1} = a(x_t) + \theta^T b(x_t) + w_t, \quad w_t\sim\mathcal{N}(0,\sigma^2),
\end{align}
where $\theta$ and $\sigma^2$ are unknown, are $\frac{1}{T}\sum_t (x_t-a(x_{t-1}))(x_t-a(x_{t-1}))^T$, $\frac{1}{T}\sum_t b(x_{t-1})(x_t-a(x_{t-1}))^T$\\and $\frac{1}{T}\sum_t b(x_{t-1})b(x_{t-1})^T$, and $x_0$ for the initial value. The maximization problem can be solved analytically.

\subsection*{Experiment 6.3--Hyperparameter estimation in infinite factorial dynamical models}
The exact setup is a replica of \cite{ValeraRSP:2015}, to which we refer for details. We use $\gamma_k = k^{-0.7}$, but let the PMCMC run for 500 iterations (which, by a very quick look at the trace of PGAS, appears to be a rough estimate of the burn-in period) before starting PSAEM. The initial value of $\eta$ are the ones chosen by \cite{ValeraRSP:2015}. The sufficient statistics for $M$ number of Beta random variables $\theta_m$ is $M$, $\sum_{m=1}^M\log(\theta_m)$ and $\sum_{m=1}^M\log(1-\theta_m)$. The maximization problem lacks an analytical solution, and an off-the-shelf numerical optimization routine (\texttt{fmincon} in Matlab) was applied to solve the maximization problem.

\subsection*{Experiment 6.4--Hyperparameter estimation in Gaussian process state-space models}
The true functions in the example are $x_{t+1}\sim \mathcal{N}\left(-7\arctan(\frac{x_t}{3})\cos(\frac{x_t}{3})\exp(-\frac{|x_t|}{10}),1\right)$ and $x_{t+1}\sim \mathcal{N}\left(-7\sin(\frac{x_t}{10},1\right)$, respectively.

In the approximate GP-SSM model used, the unknown function $f$ is approximated as a finite basis function expansion, whose coefficients $\theta$ (column vector) have a certain multivariate zero mean Gaussian prior distribution with a variance depending on $\eta$ (see \cite{SSS+:2016} for details). Thus, the sufficient statistics is $\theta\theta^T$, and the maximization problem to solve is $\arg\max_\eta -\frac{1}{2}\text{Tr}(\theta\theta^T V^{-1}_\eta) - \frac{1}{2}\log\det(V_\eta)$ (where $V_\eta$ follows from the choice of covariance function, see again \cite{SSS+:2016}), which requires a numerical approach.

%
%

%


\section*{Acknowledgment}
We would like to thank Dr. Johan Alenl\"{o}v for providing the implementation of the PaRIS and online-EM algorithm. This research was financially supported by the Swedish Foundation for Strategic Research (SSF) via the projects \emph{ASSEMBLE} (contract number: RIT15-0012) and 
\emph{Probabilistic Modeling and Inference for Machine Learning} (contract number: ICA16-0015),
and by the Swedish Research Council via the project
\emph{Learning of Large-Scale Probabilistic Dynamical Models} (contract number: 2016-04278).



\bibliographystyle{IEEEtran}
\bibliography{references}
\end{document}

%% file: fig/GPi.tex
%
%
%
\definecolor{mycolor1}{rgb}{0.98039,0.65098,0.07451}%
\definecolor{mycolor2}{rgb}{1.00000,0.38431,0.00392}%
\definecolor{mycolor3}{rgb}{0.12941,0.46275,1.00000}%
\definecolor{mycolor4}{rgb}{0.40784,0.55686,0.14902}%
\definecolor{mycolor5}{rgb}{0.68627,0.21961,0.00000}%
\begin{tikzpicture}

\begin{axis}[%
width=.8\linewidth,
height=.54\linewidth,
scale only axis,
xmin=-30,
xmax=30,
xtick={-20, -10,   0,  10,  20},
xlabel={$x_t$},
ymin=-11,
ymax=25,
ytick={-10,  -5,   0,   5,  10},
ylabel={$x_{t+1}$},
legend style={legend cell align=left, align=left, font=\tiny}
]
\addplot [color=mycolor1, dotted, line width=1.6pt]
  table[row sep=crcr]{%
-30	0.98784005641907\\
-29.5	1.33295853152719\\
-29	1.67474530449788\\
-28.5	2.01234608639781\\
-28	2.34491705109134\\
-27.5	2.67162694436632\\
-27	2.99165916163681\\
-26.5	3.30421378902918\\
-26	3.60850960275025\\
-25.5	3.90378602173992\\
-25	4.1893050087277\\
-24.5	4.46435291494153\\
-24	4.72824226385806\\
-23.5	4.98031346953591\\
-23	5.21993648523704\\
-22.5	5.44651237821545\\
-22	5.65947482673713\\
-21.5	5.85829153558948\\
-21	6.04246556654212\\
-20.5	6.21153658043363\\
-20	6.36508198777977\\
-19.5	6.50271800502709\\
-19	6.6241006138119\\
-18.5	6.7289264208271\\
-18	6.81693341614737\\
-17.5	6.88790162811756\\
-17	6.94165367316728\\
-16.5	6.97805519917743\\
-16	6.99701522129054\\
-15.5	6.9984863493255\\
-15	6.98246490622838\\
-14.5	6.94899093726312\\
-14	6.89814810991922\\
-13.5	6.83006350478661\\
-13	6.74490729792035\\
-12.5	6.6428923354891\\
-12	6.52427360177058\\
-11.5	6.38934758182365\\
-11	6.23845152043005\\
-10.5	6.07196257915812\\
-10	5.89029689365528\\
-9.5	5.69390853352562\\
-9	5.48328836739238\\
-8.5	5.25896283598205\\
-8	5.02149263629666\\
-7.5	4.77147132016334\\
-7	4.50952381066384\\
-6.5	4.23630484015228\\
-6	3.95249731376525\\
-5.5	3.65881060251461\\
-5	3.35597877022942\\
-4.5	3.04475873877861\\
-4	2.72592839616055\\
-3.5	2.40028465218816\\
-3	2.06864144662938\\
-2.5	1.73182771478166\\
-2	1.39068531556543\\
-1.5	1.04606692731519\\
-1	0.698833916527797\\
-0.5	0.349854184894748\\
0	-0\\
0.5	-0.349854184894748\\
1	-0.698833916527797\\
1.5	-1.04606692731519\\
2	-1.39068531556543\\
2.5	-1.73182771478166\\
3	-2.06864144662938\\
3.5	-2.40028465218816\\
4	-2.72592839616055\\
4.5	-3.04475873877861\\
5	-3.35597877022942\\
5.5	-3.65881060251461\\
6	-3.95249731376525\\
6.5	-4.23630484015228\\
7	-4.50952381066384\\
7.5	-4.77147132016334\\
8	-5.02149263629666\\
8.5	-5.25896283598205\\
9	-5.48328836739238\\
9.5	-5.69390853352562\\
10	-5.89029689365528\\
10.5	-6.07196257915812\\
11	-6.23845152043005\\
11.5	-6.38934758182365\\
12	-6.52427360177058\\
12.5	-6.6428923354891\\
13	-6.74490729792035\\
13.5	-6.83006350478661\\
14	-6.89814810991922\\
14.5	-6.94899093726312\\
15	-6.98246490622838\\
15.5	-6.9984863493255\\
16	-6.99701522129054\\
16.5	-6.97805519917743\\
17	-6.94165367316728\\
17.5	-6.88790162811756\\
18	-6.81693341614737\\
18.5	-6.7289264208271\\
19	-6.6241006138119\\
19.5	-6.50271800502709\\
20	-6.36508198777977\\
20.5	-6.21153658043363\\
21	-6.04246556654212\\
21.5	-5.85829153558948\\
22	-5.65947482673713\\
22.5	-5.44651237821545\\
23	-5.21993648523704\\
23.5	-4.98031346953591\\
24	-4.72824226385806\\
24.5	-4.46435291494153\\
25	-4.1893050087277\\
25.5	-3.90378602173992\\
26	-3.60850960275025\\
26.5	-3.30421378902918\\
27	-2.99165916163681\\
27.5	-2.67162694436632\\
28	-2.34491705109134\\
28.5	-2.01234608639781\\
29	-1.67474530449788\\
29.5	-1.33295853152719\\
30	-0.98784005641907\\
};
\addlegendentry{True function}

\addplot [color=mycolor2, only marks, mark size=1.6pt, mark=*, mark options={solid, mycolor2}]
  table[row sep=crcr]{%
-2.05236173080603	1.49259801580749\\
13.3042584354628	-7.12095862146838\\
-14.7054915943053	7.89359226207122\\
-3.202538122944	4.00085293095931\\
-4.45465946911866	2.38036307762447\\
-3.34628558695503	1.76368051165673\\
-3.82312713308299	3.2235928449717\\
5.00739510346936	-3.70615263016474\\
-5.67476709492417	3.07695502589339\\
8.94138124256407	-6.93936291539811\\
-15.4582326116228	6.96936421284426\\
14.9725713108625	-6.7221805249935\\
-21.8162277723374	5.93396030222061\\
14.6927017805659	-8.23399797222974\\
-10.6618933355631	6.72185944183524\\
8.38999383637027	-7.06540699746256\\
-26.7195940902904	3.45088504902136\\
-9.24982887736408	6.0237033154195\\
17.7754159700497	-5.55140120079034\\
14.7402006467073	-8.75365754056831\\
-11.5052299473253	6.06263362225385\\
12.099218080512	-5.43728589539345\\
12.3752330370316	-5.71610625847758\\
12.0204770047839	-5.07706227519639\\
-23.7282880205027	5.51255562815929\\
-4.99851496424658	2.75983003264215\\
-1.4139904473258	0.63750029338758\\
14.6591231567657	-8.59811007285468\\
-22.2755799826158	5.41490892023829\\
};
\addlegendentry{State-space samples (not part of training data)}

\addplot [color=mycolor3, line width=1.2pt]
  table[row sep=crcr]{%
-30	0.0472516432348959\\
-29.5	-0.173472321496918\\
-29	-0.276039134282076\\
-28.5	-0.193059518621198\\
-28	0.0897362982701153\\
-27.5	0.529608581230065\\
-27	1.04428984804387\\
-26.5	1.5445624102268\\
-26	1.96930331909518\\
-25.5	2.3085867147114\\
-25	2.60429214573133\\
-24.5	2.92612384341815\\
-24	3.33105912161329\\
-23.5	3.82214249043383\\
-23	4.32484548895909\\
-22.5	4.6943995757104\\
-22	4.75669616164827\\
-21.5	4.37211446030099\\
-21	3.50088527230408\\
-20.5	2.24478403638004\\
-20	0.845322831176064\\
-19.5	-0.367846717554245\\
-19	-1.06712051962003\\
-18.5	-1.02919763667005\\
-18	-0.215590745983632\\
-17.5	1.19827644637343\\
-17	2.859980077485\\
-16.5	4.33614685957252\\
-16	5.24010068221638\\
-15.5	5.35104863704738\\
-15	4.68530537616035\\
-14.5	3.49420248687566\\
-14	2.18680372260234\\
-13.5	1.20060015623195\\
-13	0.86138533330794\\
-12.5	1.27807239148112\\
-12	2.30739240419899\\
-11.5	3.60059739128603\\
-11	4.71710438410968\\
-10.5	5.26770363124847\\
-10	5.04016189686604\\
-9.5	4.06614346864862\\
-9	2.60823601212888\\
-8.5	1.07266055001896\\
-8	-0.122211234089765\\
-7.5	-0.677412670962224\\
-7	-0.501997224359921\\
-6.5	0.268797286294133\\
-6	1.32906372792661\\
-5.5	2.31006370660333\\
-5	2.90247949354634\\
-4.5	2.9519757624515\\
-4	2.49771520524828\\
-3.5	1.74397604131179\\
-3	0.977939016202828\\
-2.5	0.464194575609786\\
-2	0.352905996536881\\
-1.5	0.632207119812767\\
-1	1.13905134683549\\
-0.5	1.62233444458393\\
0	1.83471418916836\\
0.5	1.62081215593866\\
1	0.971899572747444\\
1.5	0.0294128612681859\\
2	-0.962770819917439\\
2.5	-1.74043159574008\\
3	-2.10778437472472\\
3.5	-2.00373206459369\\
4	-1.52427808593321\\
4.5	-0.894185874976644\\
5	-0.397134197536262\\
5.5	-0.286472867845979\\
6	-0.704420441967661\\
6.5	-1.63466515216978\\
7	-2.90296061152065\\
7.5	-4.22573537535871\\
8	-5.29235542471794\\
8.5	-5.85669532282393\\
9	-5.81092097485742\\
9.5	-5.2195741708811\\
10	-4.30361148568372\\
10.5	-3.37861964648629\\
11	-2.76479246467048\\
11.5	-2.69453973759905\\
12	-3.24438056850543\\
12.5	-4.31078641425679\\
13	-5.63688906022621\\
13.5	-6.88218724840854\\
14	-7.71495336222299\\
14.5	-7.90067860982445\\
15	-7.3615100957923\\
15.5	-6.19069841234871\\
16	-4.6197531030593\\
16.5	-2.95002014582954\\
17	-1.47039400187778\\
17.5	-0.385828438346098\\
18	0.223497569409737\\
18.5	0.40324806323511\\
19	0.290086250478061\\
19.5	0.0556255517327768\\
20	-0.150832629310109\\
20.5	-0.242487571810089\\
21	-0.20752256560246\\
21.5	-0.0938117763182729\\
22	0.0246422942541737\\
22.5	0.0847402742786143\\
23	0.0605282322131571\\
23.5	-0.0299054806735601\\
24	-0.137107170627377\\
24.5	-0.205121857311018\\
25	-0.197093796941073\\
25.5	-0.110922922775521\\
26	0.0207739553170785\\
26.5	0.145314135478482\\
27	0.212245883251402\\
27.5	0.193683337515803\\
28	0.0945157157784194\\
28.5	-0.0510026509440259\\
29	-0.19399598478139\\
29.5	-0.290312173142148\\
30	-0.316365895491568\\
};
\addlegendentry{GP mean with fix lengthscale 0.1}

\addplot [color=mycolor4, line width=1.2pt]
  table[row sep=crcr]{%
-30	1.13649796458495\\
-29.5	1.25915901456793\\
-29	1.44461968177983\\
-28.5	1.73914575503477\\
-28	2.17613406157604\\
-27.5	2.763938498189\\
-27	3.47894898602419\\
-26.5	4.26636438513217\\
-26	5.04899566059768\\
-25.5	5.74209516592996\\
-25	6.27037991549736\\
-24.5	6.58272549494685\\
-24	6.66072241473833\\
-23.5	6.51920445573413\\
-23	6.19934937454598\\
-22.5	5.75716895979393\\
-22	5.2513825512177\\
-21.5	4.73440541066054\\
-21	4.2486032476693\\
-20.5	3.82767882366344\\
-20	3.50094782561277\\
-19.5	3.29715149474604\\
-19	3.24478885232108\\
-18.5	3.36761950439457\\
-18	3.67634579254253\\
-17.5	4.15958972569488\\
-17	4.77825323951425\\
-16.5	5.46671150789389\\
-16	6.1421665021223\\
-15.5	6.72059392566726\\
-15	7.13510101026619\\
-14.5	7.35118028989956\\
-14	7.3738760279941\\
-13.5	7.2441989827413\\
-13	7.02550167596143\\
-12.5	6.78382088661446\\
-12	6.56825729395565\\
-11.5	6.39754581381891\\
-11	6.25700102225224\\
-10.5	6.10663830646072\\
-10	5.89760670443707\\
-9.5	5.59136124075423\\
-9	5.17516345486154\\
-8.5	4.66880200666097\\
-8	4.12041423126853\\
-7.5	3.59294069002373\\
-7	3.14582336888813\\
-6.5	2.81804141544791\\
-6	2.61799874399862\\
-5.5	2.52337104718815\\
-5	2.49063035113033\\
-4.5	2.4707492274305\\
-4	2.42561615819724\\
-3.5	2.3395992813374\\
-3	2.22246747333891\\
-2.5	2.10288285191306\\
-2	2.01488797590043\\
-1.5	1.98214949395839\\
-1	2.00542488367136\\
-0.5	2.05758504213879\\
0	2.08796175049548\\
0.5	2.03467172727002\\
1	1.84094431585979\\
1.5	1.47021561966742\\
2	0.915252507348881\\
2.5	0.198646596660611\\
3	-0.635044571207504\\
3.5	-1.53243721318359\\
4	-2.44497240585838\\
4.5	-3.33829487275327\\
5	-4.19425981448402\\
5.5	-5.00528079560588\\
6	-5.76369949443327\\
6.5	-6.45092569301708\\
7	-7.03148479098623\\
7.5	-7.4555962057049\\
8	-7.67091939963999\\
8.5	-7.64061148842438\\
9	-7.36203924269863\\
9.5	-6.87941036631865\\
10	-6.28473979566666\\
10.5	-5.70472738722588\\
11	-5.27538684169343\\
11.5	-5.1103071294572\\
12	-5.27092551218743\\
12.5	-5.74726407795153\\
13	-6.4550936322972\\
13.5	-7.25111678390404\\
14	-7.96279619379509\\
14.5	-8.42539169015294\\
15	-8.51683708414114\\
15.5	-8.18184532802224\\
16	-7.4398011340852\\
16.5	-6.37556754498929\\
17	-5.11688516669037\\
17.5	-3.80523495399849\\
18	-2.56801680201091\\
18.5	-1.49855766466747\\
19	-0.647427748502811\\
19.5	-0.0249145594434852\\
20	0.388532809161561\\
20.5	0.628654366918428\\
21	0.735117106401848\\
21.5	0.743636505092753\\
22	0.683228024875445\\
22.5	0.577322170997464\\
23	0.446383639835902\\
23.5	0.309751144946028\\
24	0.185433157289697\\
24.5	0.0879935848028746\\
25	0.0257962641810821\\
25.5	-0.000715646629954048\\
26	0.00153489493263281\\
26.5	0.0212986236355439\\
27	0.0475925105894865\\
27.5	0.0741585965266367\\
28	0.102051102935987\\
28.5	0.139274733854366\\
29	0.197490955924566\\
29.5	0.286896157703368\\
30	0.411040740632881\\
};
\addlegendentry{GP mean with estimated lengthscale 2.52}

\end{axis}

\end{tikzpicture}%

%% file: fig/GPii.tex
%
%
%
\definecolor{mycolor1}{rgb}{0.98039,0.65098,0.07451}%
\definecolor{mycolor2}{rgb}{1.00000,0.38431,0.00392}%
\definecolor{mycolor3}{rgb}{0.12941,0.46275,1.00000}%
\definecolor{mycolor4}{rgb}{0.40784,0.55686,0.14902}%
\begin{tikzpicture}

\begin{axis}[%
width=.8\linewidth,
height=.54\linewidth,
scale only axis,
xmin=-30,
xmax=30,
xtick={-20, -10,   0,  10,  20},
xlabel={$x_t$},
ymin=-11,
ymax=25,
ytick={-10,  -5,   0,   5,  10},
ylabel={$x_{t+1}$},
legend style={legend cell align=left, align=left, font=\tiny}
]
\addplot [color=mycolor1, dotted, line width=1.6pt]
  table[row sep=crcr]{%
-30	-0.430193599656171\\
-29.5	-0.494063510428965\\
-29	-0.548851061654354\\
-28.5	-0.59188726890538\\
-28	-0.620603488448169\\
-27.5	-0.632623782715429\\
-27	-0.625860643998084\\
-26.5	-0.598611345093231\\
-26	-0.549651879010202\\
-25.5	-0.478325208803063\\
-25	-0.384620387906593\\
-24.5	-0.269239044733576\\
-24	-0.133645764917372\\
-23.5	0.0199009394821463\\
-23	0.188340104844112\\
-22.5	0.367824858369757\\
-22	0.553766115318521\\
-21.5	0.740904857273946\\
-21	0.923413639254991\\
-20.5	1.09502709035438\\
-20	1.24920020177769\\
-19.5	1.37929215938918\\
-19	1.47877240126066\\
-18.5	1.5414444944155\\
-18	1.5616823625872\\
-17.5	1.53467239475765\\
-17	1.45665406075609\\
-16.5	1.32515089434946\\
-16	1.1391831147117\\
-15.5	0.899452780870511\\
-15	0.608492244481777\\
-14.5	0.270766813181762\\
-14	-0.107277017320394\\
-13.5	-0.517225644274164\\
-13	-0.948781554209161\\
-12.5	-1.38992104251961\\
-12	-1.82713017317303\\
-11.5	-2.24572073866119\\
-11	-2.63022580420891\\
-10.5	-2.96487208722392\\
-10	-3.23412401263325\\
-9.5	-3.42329189470448\\
-9	-3.51919444191219\\
-8.5	-3.5108637958098\\
-8	-3.39027974157793\\
-7.5	-3.15311870783333\\
-7	-2.79950280875394\\
-6.5	-2.33473445315714\\
-6	-1.77000262265526\\
-5.5	-1.12304674470391\\
-5	-0.418760461004858\\
-4.5	0.310295710953489\\
-4	1.02354032624706\\
-3.5	1.67233025434571\\
-3	2.20057613010052\\
-2.5	2.54672035376849\\
-2	2.64836988979338\\
-1.5	2.45148859527548\\
-1	1.92575004887922\\
-0.5	1.08442223606065\\
0	-0\\
0.5	-1.08442223606065\\
1	-1.92575004887922\\
1.5	-2.45148859527548\\
2	-2.64836988979338\\
2.5	-2.54672035376849\\
3	-2.20057613010052\\
3.5	-1.67233025434571\\
4	-1.02354032624706\\
4.5	-0.310295710953489\\
5	0.418760461004858\\
5.5	1.12304674470391\\
6	1.77000262265526\\
6.5	2.33473445315714\\
7	2.79950280875394\\
7.5	3.15311870783333\\
8	3.39027974157793\\
8.5	3.5108637958098\\
9	3.51919444191219\\
9.5	3.42329189470448\\
10	3.23412401263325\\
10.5	2.96487208722392\\
11	2.63022580420891\\
11.5	2.24572073866119\\
12	1.82713017317303\\
12.5	1.38992104251961\\
13	0.948781554209161\\
13.5	0.517225644274164\\
14	0.107277017320394\\
14.5	-0.270766813181762\\
15	-0.608492244481777\\
15.5	-0.899452780870511\\
16	-1.1391831147117\\
16.5	-1.32515089434946\\
17	-1.45665406075609\\
17.5	-1.53467239475765\\
18	-1.5616823625872\\
18.5	-1.5414444944155\\
19	-1.47877240126066\\
19.5	-1.37929215938918\\
20	-1.24920020177769\\
20.5	-1.09502709035438\\
21	-0.923413639254991\\
21.5	-0.740904857273946\\
22	-0.553766115318521\\
22.5	-0.367824858369757\\
23	-0.188340104844112\\
23.5	-0.0199009394821463\\
24	0.133645764917372\\
24.5	0.269239044733576\\
25	0.384620387906593\\
25.5	0.478325208803063\\
26	0.549651879010202\\
26.5	0.598611345093231\\
27	0.625860643998084\\
27.5	0.632623782715429\\
28	0.620603488448169\\
28.5	0.59188726890538\\
29	0.548851061654354\\
29.5	0.494063510428965\\
30	0.430193599656171\\
};
\addlegendentry{True function}

\addplot [color=mycolor2, only marks, mark size=1.6pt, mark=*, mark options={solid, mycolor2}]
  table[row sep=crcr]{%
-2.05236173080603	2.71692626775464\\
14.52858668741	-0.613442497293814\\
-8.19797547013073	-2.52308078437296\\
-13.6192111693882	1.38009339492089\\
-7.07541900515708	-3.49593955338387\\
-9.22258821796338	-4.02402861936267\\
-9.61083626410239	-2.77684841760674\\
-0.993046159109072	1.57048578255522\\
-0.398128682204212	0.194131833863468\\
6.05855805053416	0.359121743448547\\
-8.15974795277614	-3.46980757724768\\
4.53339952077055	-0.00269495878508863\\
-15.096742206129	0.868343874253835\\
9.62708535259916	2.11351791580385\\
-0.314377447529549	1.29868230718338\\
2.96681670171841	-4.08747009106599\\
-23.7416571838938	0.22659261503364\\
-12.4741213113518	-0.979338280885642\\
10.7723743737445	4.08923129972867\\
24.3808331472263	-1.54772595543605\\
-4.299298362193	0.272837027915783\\
6.30942148617393	3.24229271531929\\
21.0548116477443	-0.00509156625815166\\
17.7314916970033	-0.100926367469341\\
-18.7521521127757	2.16063094097488\\
-8.350439651431	-4.08207215211546\\
-8.25589263208341	-3.8153781137893\\
10.2062447495888	1.49537582220386\\
-12.1820940875573	-1.79817094514013\\
};
\addlegendentry{State-space samples (not part of training data)}

\addplot [color=mycolor3, line width=1.2pt]
  table[row sep=crcr]{%
-30	-2.76973766733401\\
-29.5	-2.82711977112353\\
-29	-2.82976033394182\\
-28.5	-2.77650349418011\\
-28	-2.66724856380131\\
-27.5	-2.50302662483524\\
-27	-2.28605839498931\\
-26.5	-2.01978828663343\\
-26	-1.70888988670757\\
-25.5	-1.35923867915947\\
-25	-0.977848713446238\\
-24.5	-0.572771073917656\\
-24	-0.152953391164098\\
-23.5	0.27193879196583\\
-23	0.691736293321453\\
-22.5	1.09601195595172\\
-22	1.474351750907\\
-21.5	1.81664378141324\\
-21	2.11337627671413\\
-20.5	2.35593458011885\\
-20	2.53688641597049\\
-19.5	2.65024442379939\\
-19	2.69169511700784\\
-18.5	2.65878407984969\\
-18	2.55104835946633\\
-17.5	2.37008861485602\\
-17	2.1195756036956\\
-16.5	1.80518795015654\\
-16	1.43448075132069\\
-15.5	1.01668733885586\\
-15	0.562459296400785\\
-14.5	0.0835525149088658\\
-14	-0.407530479618185\\
-13.5	-0.897929591237799\\
-13	-1.37483363921984\\
-12.5	-1.82588920738572\\
-12	-2.23959634306836\\
-11.5	-2.605675798458\\
-11	-2.9153933912025\\
-10.5	-3.16182860009887\\
-10	-3.34007665095422\\
-9.5	-3.44737600165171\\
-9	-3.48315619106465\\
-8.5	-3.44900433819474\\
-8	-3.34855201421908\\
-7.5	-3.18728760090338\\
-7	-2.97230243351841\\
-6.5	-2.71198185227073\\
-6	-2.41565461648111\\
-5.5	-2.0932158565907\\
-5	-1.75473976604979\\
-4.5	-1.41009851755141\\
-4	-1.06860341181837\\
-3.5	-0.738683055608885\\
-3	-0.427611478502946\\
-2.5	-0.141296628443117\\
-2	0.115863244389046\\
-1.5	0.341051052884821\\
-1	0.533029329674942\\
-0.5	0.692068213294318\\
0	0.819808724448927\\
0.5	0.919069986712947\\
1	0.99361147314506\\
1.5	1.04786320734845\\
2	1.08663803075344\\
2.5	1.11484051775661\\
3	1.13718686364026\\
3.5	1.15794911076731\\
4	1.1807354759245\\
4.5	1.20831638787189\\
5	1.24250325871791\\
5.5	1.28408413639514\\
6	1.33281737255086\\
6.5	1.38748145007292\\
7	1.44597630322878\\
7.5	1.50546897494781\\
8	1.56257441414963\\
8.5	1.61356071858868\\
9	1.65456724098942\\
9.5	1.68182372872094\\
10	1.69185905405802\\
10.5	1.68168907209743\\
11	1.64897464343967\\
11.5	1.59214277856093\\
12	1.51046607972828\\
12.5	1.40409804100824\\
13	1.27406417967239\\
13.5	1.12221127934436\\
14	0.951119104521119\\
14.5	0.763980694053716\\
15	0.564458677951188\\
15.5	0.356525934922258\\
16	0.144299293787513\\
16.5	-0.0681251146146983\\
17	-0.276826795314191\\
17.5	-0.478198999432204\\
18	-0.669064153303125\\
18.5	-0.846762252132642\\
19	-1.00920950398924\\
19.5	-1.15492619567916\\
20	-1.28303439178982\\
20.5	-1.39322758886611\\
21	-1.48571575765316\\
21.5	-1.56115026210742\\
22	-1.62053390584002\\
22.5	-1.66512180376356\\
23	-1.69631890519954\\
23.5	-1.71557981696594\\
24	-1.72431611778071\\
24.5	-1.7238156575264\\
25	-1.71517744476565\\
25.5	-1.69926469814017\\
26	-1.67667753027554\\
26.5	-1.64774560559217\\
27	-1.61254002299417\\
27.5	-1.57090267325992\\
28	-1.52249045490218\\
28.5	-1.46683103867011\\
29	-1.40338637732301\\
29.5	-1.33161988077458\\
30	-1.25106312314271\\
};
\addlegendentry{GP mean with fix lengthscale 8}

\addplot [color=mycolor4, line width=1.2pt]
  table[row sep=crcr]{%
-30	0.575817933585134\\
-29.5	0.693087859024144\\
-29	0.711730734733175\\
-28.5	0.627915133409391\\
-28	0.454019531953824\\
-27.5	0.216942619233874\\
-27	-0.0458097519301421\\
-26.5	-0.291646964455722\\
-26	-0.479164460962452\\
-25.5	-0.574192618060947\\
-25	-0.554698060875581\\
-24.5	-0.413802750151575\\
-24	-0.160491793731505\\
-23.5	0.182048639373146\\
-23	0.580146828332971\\
-22.5	0.994778803842217\\
-22	1.38727178662067\\
-21.5	1.72465859198524\\
-21	1.98384298930677\\
-20.5	2.1539511214834\\
-20	2.23655810608825\\
-19.5	2.24383868602661\\
-19	2.1950359233554\\
-18.5	2.11191504296932\\
-18	2.01402583075918\\
-17.5	1.9146117528195\\
-17	1.81787661749747\\
-16.5	1.71807367073192\\
-16	1.60056092474066\\
-15.5	1.44462601311502\\
-15	1.22758231050586\\
-14.5	0.929426347507546\\
-14	0.537259366707433\\
-13.5	0.048725802660986\\
-13	-0.526102891371587\\
-12.5	-1.16470242245707\\
-12	-1.8339190547961\\
-11.5	-2.49333272339845\\
-11	-3.09961310152963\\
-10.5	-3.61118744679566\\
-10	-3.99252734470247\\
-9.5	-4.21746950501514\\
-9	-4.2711887257295\\
-8.5	-4.15069736232734\\
-8	-3.86400475471788\\
-7.5	-3.42828270154168\\
-7	-2.86751033047995\\
-6.5	-2.21009258018799\\
-6	-1.48686180202392\\
-5.5	-0.729704046816342\\
-5	0.0291602548988557\\
-4.5	0.757418907607429\\
-4	1.4227668790177\\
-3.5	1.99326079868773\\
-3	2.43828687101427\\
-2.5	2.73035524490413\\
-2	2.84771266534088\\
-1.5	2.77751260041947\\
-1	2.51905255341311\\
-0.5	2.08643534816526\\
0	1.50997615925427\\
0.5	0.835780198641628\\
1	0.123151155413989\\
1.5	-0.560175130584007\\
2	-1.14459846877721\\
2.5	-1.56629449389318\\
3	-1.77493328721804\\
3.5	-1.7403643448373\\
4	-1.45721663348313\\
4.5	-0.946604102436247\\
5	-0.254518828592185\\
5.5	0.553022959786292\\
6	1.39752020151701\\
6.5	2.19722331709141\\
7	2.8767969263032\\
7.5	3.37605218557361\\
8	3.65652110235381\\
8.5	3.70497220937059\\
9	3.53343997176312\\
9.5	3.17588135437909\\
10	2.68209153815143\\
10.5	2.10991946199044\\
11	1.51705397544095\\
11.5	0.953665239001523\\
12	0.456983853011513\\
12.5	0.0485203336126054\\
13	-0.265860422164119\\
13.5	-0.493302734347946\\
14	-0.649491052141876\\
14.5	-0.753364693728777\\
15	-0.822121729052396\\
15.5	-0.867495130545937\\
16	-0.893977883107864\\
16.5	-0.899244117362026\\
17	-0.876544479785676\\
17.5	-0.818437052713874\\
18	-0.720931399711142\\
18.5	-0.587028576265347\\
19	-0.428753911699664\\
19.5	-0.267081544117804\\
20	-0.129582888057628\\
20.5	-0.0461120201632484\\
21	-0.0432753044917236\\
21.5	-0.138733501278479\\
22	-0.33648901030288\\
22.5	-0.624192006790037\\
23	-0.973172388534418\\
23.5	-1.34142514543083\\
24	-1.67923098365577\\
24.5	-1.93658278608036\\
25	-2.07120914679833\\
25.5	-2.05581423146793\\
26	-1.88322753503166\\
26.5	-1.5684721126225\\
27	-1.14726539882902\\
27.5	-0.671076406080778\\
28	-0.199469371506\\
28.5	0.209042592836544\\
29	0.506120914405085\\
29.5	0.661160112343074\\
30	0.66598135146842\\
};
\addlegendentry{GP mean with estimated lengthscale 0.158}

\end{axis}
\end{tikzpicture}%

%% file: psaem.bbl
\begin{thebibliography}{10}
\providecommand{\url}[1]{#1}
\csname url@samestyle\endcsname
\providecommand{\newblock}{\relax}
\providecommand{\bibinfo}[2]{#2}
\providecommand{\BIBentrySTDinterwordspacing}{\spaceskip=0pt\relax}
\providecommand{\BIBentryALTinterwordstretchfactor}{4}
\providecommand{\BIBentryALTinterwordspacing}{\spaceskip=\fontdimen2\font plus
\BIBentryALTinterwordstretchfactor\fontdimen3\font minus
  \fontdimen4\font\relax}
\providecommand{\BIBforeignlanguage}[2]{{%
\expandafter\ifx\csname l@#1\endcsname\relax
\typeout{** WARNING: IEEEtran.bst: No hyphenation pattern has been}%
\typeout{** loaded for the language `#1'. Using the pattern for}%
\typeout{** the default language instead.}%
\else
\language=\csname l@#1\endcsname
\fi
#2}}
\providecommand{\BIBdecl}{\relax}
\BIBdecl

\bibitem{FrigolaCR:2014}
R.~Frigola, Y.~Chen, and C.~E. Rasmussen, ``Variational {Gaussian} process
  state-space models,'' in \emph{Advances in Neural Information Processing
  Systems ({NIPS}) 27}, Montr\'eal, Canada, 2014.

\bibitem{MattosDD:2016}
C.~L.~C. Mattos, Z.~Dai, A.~Damianou, J.~Forth, G.~A. Barreto, and N.~D.
  Lawrence, ``Recurrent {Gaussian} processes,'' in \emph{International
  Conference on Learning representations (ICLR)}, San Juan, Puerto Rico, 2016.

\bibitem{GaelTG:2009}
J.~V. Gael, Y.~Teh, and Z.~Ghahramani, ``The infinite factorial hidden {Markov}
  model,'' in \emph{Advances in Neural Information Processing Systems (NIPS)
  21}, Vancouver, Canada, 2009, pp. 1967--1704.

\bibitem{ValeraRSP:2015}
I.~Valera, F.~Ruiz, L.~Svensson, and F.~Perez-Cruz, ``Infinite factorial
  dynamical model,'' in \emph{Advances in Neural Information Processing Systems
  (NIPS) 28}, 2015, pp. 1666--1674.

\bibitem{FraccaroSPW:2016}
M.~Fraccaro, S.~K. S{\o}nderby, U.~Paquet, and O.~Winther, ``Sequential neural
  models with stochastic layers,'' in \emph{Advances in Neural Information
  Processing Systems (NIPS) 29}, Barcelona, Spain, 2016, pp. 2199--2207.

\bibitem{Digalakis1993}
V.~Digalakis, J.~R. Rohlicek, and M.~Ostendorf, ``{ML} estimation of a
  stochastic linear system with the {EM} algorithm and its application to
  speech recognition,'' \emph{IEEE Transactions on Speech and Audio
  Processing}, vol.~1, no.~4, pp. 431--442, 1993.

\bibitem{GhahramaniH:1996}
Z.~Ghahramani and G.~E. Hinton, ``Parameter estimation for linear dynamical
  systems,'' Department of Computer Science, University of Toronto, Tech. Rep.
  CRG-TR-96-2, 1996.

\bibitem{DempsterLR:1977}
A.~Dempster, N.~Laird, and D.~Rubin, ``Maximum likelihood from incomplete data
  via the {EM} algorithm,'' \emph{Journal of the {R}oyal {S}tatistical
  {S}ociety, {S}eries {B}}, vol.~39, no.~1, pp. 1--38, 1977.

\bibitem{CappeMR:2005}
O.~Capp\'e, E.~Moulines, and T.~Ryd\'en, \emph{Inference in Hidden {M}arkov
  Models}.\hskip 1em plus 0.5em minus 0.4em\relax New York, NY, USA: Springer,
  2005.

\bibitem{OlssonDCM:2008}
J.~Olsson, R.~Douc, O.~Capp{\'{e}}, and E.~Moulines, ``Sequential {M}onte
  {C}arlo smoothing with application to parameter estimation in nonlinear
  state-space models,'' \emph{Bernoulli}, vol.~14, no.~1, pp. 155--179, 2008.

\bibitem{SchonWN:2011}
T.~B. Sch\"{o}n, A.~Wills, and B.~Ninness, ``System identification of nonlinear
  state-space models,'' \emph{Automatica}, vol.~47, no.~1, pp. 39--49, 2011.

\bibitem{RobertC:2004}
C.~P. Robert and G.~Casella, \emph{{M}onte {C}arlo Statistical Methods}.\hskip
  1em plus 0.5em minus 0.4em\relax Springer, 2004.

\bibitem{Tierney:1994}
L.~Tierney, ``{M}arkov chains for exploring posterior distributions,''
  \emph{The Annals of Statistics}, vol.~22, no.~4, pp. 1701--1728, 1994.

\bibitem{DoucetJ:2011}
A.~Doucet and A.~Johansen, ``A tutorial on particle filtering and smoothing:
  Fifteen years later,'' in \emph{The Oxford Handbook of Nonlinear Filtering},
  D.~Crisan and B.~Rozovskii, Eds.\hskip 1em plus 0.5em minus 0.4em\relax
  Oxford, UK: Oxford University Press, 2011, pp. 656--704.

\bibitem{DoucetFG:2001}
A.~Doucet, N.~de~Freitas, and N.~Gordon, Eds., \emph{Sequential {M}onte {C}arlo
  Methods in Practice}.\hskip 1em plus 0.5em minus 0.4em\relax New York, USA:
  Springer Verlag, 2001.

\bibitem{Wu:1983}
C.~F.~J. Wu, ``On the convergence properties of the {EM} algorithm,'' \emph{The
  Annals of Statistics}, vol.~11, no.~1, pp. 95--103, 1983.

\bibitem{AndrieuDH:2010}
C.~Andrieu, A.~Doucet, and R.~Holenstein, ``Particle {M}arkov chain {M}onte
  {C}arlo methods,'' \emph{Journal of the Royal Statistical Society: Series B},
  vol.~72, no.~3, pp. 269--342, 2010.

\bibitem{DelyonLM:1999}
B.~Delyon, M.~Lavielle, and E.~Moulines, ``Convergence of a stochastic
  approximation version of the {EM} algorithm,'' \emph{The Annals of
  Statistics}, vol.~27, no.~1, pp. 94--128, 1999.

\bibitem{GR:1999}
Z.~Ghahramani and S.~T. Roweis, ``Learning nonlinear dynamical systems using an
  {EM} algorithm,'' in \emph{Advances in Neural Information Processing Systems
  (NIPS) 11}, Denver, CO, USA, Nov. 1998, pp. 431--437.

\bibitem{DL:2013}
M.~Delattre and M.~Lavielle, ``Coupling the {SAEM} algorithm and the extended
  {Kalman} filter for maximum likelihood estimation in mixed-effects diffusion
  models,'' \emph{Statistics and Its Interface}, vol.~6, pp. 519--532, 2013.

\bibitem{UmenbergerWM:2018}
J.~Umenberger, J.~W{\aa}gberg, I.~Manchester, and T.~B. Sch\"on, ``Maximum
  likelihood identification of stable linear dynamical systems,''
  \emph{Automatica}, 2018, forthcoming, provisionally accepted,.

\bibitem{AndrieuD:2003}
C.~{Andrieu} and A.~{Doucet}, ``Online expectation-maximization type algorithms
  for parameter estimation in general state space models,'' in
  \emph{Proceedings of the 28th {IEEE} International Conference on Acoustics,
  Speech and Signal Processing ({ICASSP})}, Hong Kong, China, Apr. 2003, pp.
  VI--69--VI--74.

\bibitem{AndrieuDT:2005}
C.~Andrieu, A.~Doucet, and V.~B. Tadi\'c, ``On-line parameter estimation in
  general state-space models,'' in \emph{Proceedings of the 44th {IEEE}
  Conference on Decision and Control ({CDC})}, Seville, Spain, Dec. 2005, pp.
  332--337.

\bibitem{OW:2015}
J.~Olsson and J.~Westerborn, ``An efficient particle-based online {EM}
  algorithm for general state-space models,'' in \emph{Proceedings of the 17th
  IFAC Symposium on System Identification (SYSID)}, Beijing, China, Oct. 2015,
  pp. 963--968.

\bibitem{KuhnL:2004}
E.~Kuhn and M.~Lavielle, ``Coupling a stochastic approximation version of {EM}
  with an {MCMC} procedure,'' \emph{ESAIM: Probability and Statistics}, vol.~8,
  pp. 115--131, 2004.

\bibitem{DonnetS:2011}
S.~Donnet and A.~Samson, ``{EM} algorithm coupled with particle filter for
  maximum likelihood parameter estimation of stochastic differential
  mixed-effects models,'' Universit\'e Paris Descartes, MAP5, Tech. Rep.
  hal-00519576, v2, 2011.

\bibitem{AndrieuV:2014}
C.~Andrieu and M.~Vihola, ``Markovian stochastic approximation with expanding
  projections,'' \emph{Bernoulli}, vol.~20, no.~2, pp. 545--585, 2014.

\bibitem{LindstenJS:2014}
F.~Lindsten, M.~I. Jordan, and T.~B. Sch\"on, ``Particle {G}ibbs with ancestor
  sampling,'' \emph{Journal of Machine Learning Research}, vol.~15, pp.
  2145--2184, 2014.

\bibitem{Lindsten:2013}
F.~Lindsten, ``An efficient stochastic approximation {EM} algorithm using
  conditional particle filters,'' in \emph{Proceedings of the 38th {IEEE}
  International Conference on Acoustics, Speech and Signal Processing
  ({ICASSP})}, Vancouver, Canada, 2013.

\bibitem{SS:2017}
A.~Svensson and T.~B. Sch\"{o}n, ``A flexible state space model for learning
  nonlinear dynamical systems,'' \emph{Automatica}, vol.~80, pp. 189--199,
  2017.

\bibitem{SvenssonLS:2014}
A.~Svensson, F.~Lindsten, and T.~B. Sch\"on, ``Identification of jump {M}arkov
  linear models using particle filters,'' in \emph{Proceedings of the 53rd
  {IEEE} Conference on Decision and Control ({CDC})}, Los Angeles, USA, 2014.

\bibitem{GongZS:2017}
M.~Gong, K.~Zhang, B.~Sch\"olkopf, C.~Glymour, and D.~Tao, ``Causal discovery
  from temporally aggregated time series,'' in \emph{Proceedings of the
  Conference on Uncertainty in Artificial Intelligence}, Sydney, Australia,
  2017.

\bibitem{SingorBA:2017}
S.~N. Singor, A.~Boer, J.~S.~C. Alberts, and C.~W. Oosterlee, ``On the
  modelling of nested risk-neutral stochastic processes with applications in
  insurance,'' \emph{Applied Mathematical Finance}, vol.~24, no.~2, pp.
  302--336, 2017.

\bibitem{SSS+:2016}
A.~Svensson, A.~Solin, S.~S{\"a}rkk{\"a}, and T.~B. Sch\"{o}n,
  ``Computationally efficient {Bayesian} learning of {Gaussian} process state
  space models,'' in \emph{Proceedings of the 19th International Conference on
  Artificial Intelligence and Statistics (AISTATS)}, Cadiz, Spain, 2016, pp.
  213--221.

\bibitem{LindermanSA:2014}
S.~Linderman, C.~H. Stock, and R.~P. Adams, ``A framework for studying synaptic
  plasticity with neural spike train data,'' in \emph{Advances in Neural
  Information Processing Systems ({NIPS}) 27}, Montr\'eal, Canada, 2014.

\bibitem{MeentHMW:2015}
J.-W. van~de Meent, Y.~Hongseok, V.~Mansinghka, and F.~Wood, ``Particle {G}ibbs
  with ancestor sampling for probabilistic programs,'' in \emph{Proceedings of
  the 18th International Conference on Artificial Intelligence and Statistics
  ({AISTATS})}, San Diego, CA, USA, 2015.

\bibitem{MarcosCBD:2015}
M.~Marcos, F.~M. Calafat, A.~Berihuete, and S.~Dangendorf, ``Long-term
  variations in global sea level extremes,'' \emph{Journal of Geophysical
  Research}, vol. 120, no.~12, pp. 8115--8134, 2015.

\bibitem{Whiteley:2010}
N.~Whiteley, ``Discussion on {P}article {M}arkov chain {M}onte {C}arlo
  methods,'' \emph{Journal of the Royal Statistical Society: Series {B}},
  vol.~72, no.~3, pp. 306--307, 2010.

\bibitem{LindstenS:2012}
F.~Lindsten and T.~B. Sch\"{o}n, ``On the use of backward simulation in the
  particle {G}ibbs sampler,'' in \emph{Proceedings of the 37th {IEEE}
  International Conference on Acoustics, Speech and Signal Processing
  ({ICASSP})}, Kyoto, Japan, Mar. 2012.

\bibitem{SinghLM:2015}
S.~S. Singh, F.~Lindsten, and E.~Moulines, ``Blocking strategies and stability
  of particle {G}ibbs samplers,'' \emph{Biometrika}, vol. 104, no.~4, pp.
  953--969, 2017.

\bibitem{ChopinS:2015}
N.~Chopin and S.~S. Singh, ``On particle {G}ibbs sampling,'' \emph{Bernoulli},
  vol.~21, no.~3, pp. 1855--1883, 2015.

\bibitem{LindstenDM:2015}
F.~Lindsten, R.~Douc, and E.~Moulines, ``Uniform ergodicity of the particle
  {G}ibbs sampler,'' \emph{Scandinavian Journal of Statistics}, vol.~42, no.~3,
  pp. 775--797, 2015.

\bibitem{AndrieuLV:2018}
C.~Andrieu, A.~Lee, and M.~Vihola, ``Uniform ergodicity of the iterated
  conditional {SMC} and geometric ergodicity of particle {G}ibbs samplers,''
  \emph{Bernoulli}, vol.~24, no.~2, pp. 842--872, 2018.

\bibitem{DelMoralKP:2014}
P.~Del~Moral, R.~Kohn, and F.~Patras, ``On particle {G}ibbs {M}arkov chain
  {M}onte {C}arlo models,'' arXiv:1404.5733, 2014.

\bibitem{WeiT:1990}
G.~C.~G. Wei and M.~A. Tanner, ``A {M}onte {C}arlo implementation of the {EM}
  algorithm and the poor man's data augmentation algorithms,'' \emph{Journal of
  the American Statistical Association}, vol.~85, no. 411, pp. 699--704, 1990.

\bibitem{DieboltI:1996}
J.~Diebolt and E.~H.~S. Ip, ``{Stochastic EM}: method and application,'' in
  \emph{Markov Chain Monte Carlo in Practice}, W.~R. Gilks, S.~Richardson, and
  D.~J. Spiegelhalter, Eds.\hskip 1em plus 0.5em minus 0.4em\relax Boca Raton,
  FL, USA: Chapman \& Hall/CRC, 1996, pp. 259--274.

\bibitem{FortM:2003}
G.~Fort and E.~Moulines, ``Convergence of the {M}onte {C}arlo expectation
  maximization for curved exponential families,'' \emph{The Annals of
  Statistics}, vol.~31, no.~4, pp. 1220--1259, 2003.

\bibitem{RM:1951}
H.~Robbins and S.~Monro, ``A stochastic approximation method,'' \emph{The
  Annals of Mathematical Statistics}, vol.~22, no.~3, pp. 400--407, 1951.

\bibitem{AndrieuMP:2005}
C.~Andrieu, E.~Moulines, and P.~Priouret, ``Stability of stochastic
  approximation under verifiable conditions,'' \emph{SIAM Journal on Control
  and Optimization}, vol.~44, no.~1, pp. 283--312, 2005.

\bibitem{DelMoral:2004}
P.~Del~Moral, \emph{{Feynman}-{Kac} Formulae - Genealogical and Interacting
  Particle Systems with Applications}, ser. Probability and its
  Applications.\hskip 1em plus 0.5em minus 0.4em\relax New York, USA: Springer,
  2004.

\bibitem{Whiteley:2013}
N.~Whiteley, ``Stability properties of some particle filters,'' \emph{Annals of
  Applied Probability}, vol.~23, no.~6, pp. 2500--2537, 2013.

\bibitem{Handel:2009}
R.~v. Handel, ``Uniform time average consistency of {M}onte {C}arlo particle
  filters,'' \emph{Stochastic Processes and their Applications}, vol. 119,
  no.~11, pp. 3835--3861, 2009.

\bibitem{DonnetS:2014}
S.~Donnet and A.~Samson, ``Using {PMCMC} in {EM} algorithm for stochastic mixed
  models: theoretical and practical issues,'' \emph{Journal de la Societe
  Fran\c{c}aise de Statistique}, vol. 155, no.~1, pp. 49--72, 2014.

\bibitem{GodsillDW:2004}
S.~J. Godsill, A.~Doucet, and M.~West, ``{M}onte {C}arlo smoothing for
  nonlinear time series,'' \emph{Journal of the American Statistical
  Association}, vol.~99, no. 465, pp. 156--168, Mar. 2004.

\bibitem{OlssonW:2017}
J.~Olsson and J.~Westerborn, ``Efficient particle-based online smoothing in
  general hidden {Markov} models: the {PaRIS} algorithm,'' \emph{Bernoulli},
  vol.~23, no.~3, pp. 1951--1996, 2017.

\bibitem{SN:2017}
M.~Schoukens and J.-P. No\"{e}l, ``Three benchmarks addressing open challenges
  in nonlinear system identification,'' in \emph{Proceedings of the 20th World
  Congress of the International Federation of Automatic Control (IFAC)},
  Toulouse, France, Jul. 2017.

\bibitem{HRC+:16}
G.~Holmes, T.~Rogers \emph{et~al.}, ``Cascaded tanks benchmark: Parametric and
  nonparametric identification,'' Presentation at \textit{Workshop on Nonlinear
  System Identification Benchmarks 2016}, Vrije Universiteit Brussel, Brussels,
  Belgium, May 2016.

\bibitem{RTM:2017}
R.~Relan, K.~Tiels, A.~Marconato, and J.~Schoukens, ``An unstructured flexible
  nonlinear model for the cascaded water-tanks benchmark,'' in
  \emph{Proceedings of the 20th International Federation of Automatic Control
  World Congress (IFAC)}, 2017, pp. 454--459.

\bibitem{RasmussenW:2006}
C.~E. Rasmussen and C.~K.~I. Williams, \emph{Gaussian Processes for Machine
  Learning}.\hskip 1em plus 0.5em minus 0.4em\relax {MIT} Press, 2006.

\bibitem{FrigolaLSR:2013}
R.~Frigola, F.~Lindsten, T.~B. Sch\"on, and C.~E. Rasmussen, ``{B}ayesian
  inference and learning in {G}aussian process state-space models with particle
  {MCMC},'' in \emph{Advances in Neural Information Processing Systems ({NIPS})
  26}, Lake Tahoe, NV, USA, 2013.

\bibitem{FrigolaLSR:2014}
------, ``Identification of {Gaussian} process state-space models with particle
  stochastic approximation {EM},'' in \emph{Proceedings of the 19th
  International Federation of Automatic Control World Congress (IFAC)}, Cape
  Town, South Africa, 2014.

\bibitem{JacobLS:2017}
P.~E. Jacob, F.~Lindsten, and T.~B. Sch\"on, ``Smoothing with couplings of
  conditional particle filters,'' \emph{arXiv:1701.02002}, 2017.

\end{thebibliography}
